\documentclass{jpp}
\usepackage{graphicx}
\usepackage{epstopdf, epsfig}

\usepackage{mathrsfs}
\usepackage{amsmath}

\usepackage{amssymb}
\usepackage{color}

\newcommand\beq{\begin{equation}}
\newcommand\eeq{\end{equation}}

\newcommand{\ben}{\begin{eqnarray}}
\newcommand{\een}{\end{eqnarray}}
\newcommand{\benn}{\begin{eqnarray*}}
\newcommand{\eenn}{\end{eqnarray*}}

\newcommand{\btimes}{{\boldsymbol \times}}
\newcommand{\hatb}{{\widehat{ b}}}
\newcommand{\bhatb}{{\widehat{\boldsymbol b}}}
\newcommand{\bhatz}{{\widehat{\boldsymbol z}}}

\newcommand{\circone[1]}{\mbox{\textcircled{\small #1}}}

\title[Electron scale reduced fluid models]{Electron-scale reduced fluid models with gyroviscous effects}

\author[T. Passot,  P.L. Sulem and E. Tassi]%
       {T.\ns P\ls A\ls S\ls S\ls O\ls T$^1$
        \thanks{Email address for correspondence: passot@oca.eu}\ns, 
        P.L.\ns S\ls U\ls L\ls E\ls M$^1$ and  E.\ns T\ls A\ls S\ls S\ls I$^2$}

\affiliation{$^1$ Universit\'e C\^ote d'Azur, CNRS, Observatoire de la C\^ote d'Azur,
Laboratoire J.L. Lagrange, Boulevard de l'Observatoire, CS  34229, 06304 Nice Cedex 4, France \\[\affilskip]
$^2$ Aix Marseille Univ, Univ Toulon, CNRS, CPT, Marseille, France }

\pubyear{2010}
\volume{650}
\pagerange{119--126}
\date{?; revised ?; accepted ?. - To be entered by editorial office}

\begin{document}

\maketitle

\begin{abstract}
Reduced fluid models for collisionless plasmas including electron
inertia and finite Larmor radius corrections are derived for scales ranging
from the ion to the electron gyroradii. Based either on pressure balance
or on the incompressibility of the electron fluid, they respectively
capture kinetic Alfv\'en waves (KAWs) or whistler waves (WWs), 
and can provide suitable tools for reconnection
and turbulence studies. Both isothermal regimes and Landau fluid closures 
permitting anisotropic pressure fluctuations are considered.
For small values of the electron beta parameter $\beta_e$, a perturbative computation of the
gyroviscous force valid at scales comparable to the electron inertial length is
performed at order $O(\beta_e)$, which requires second-order
contributions in a scale expansion. 
Comparisons with kinetic theory are performed in the
linear regime. The spectrum of transverse magnetic fluctuations
for strong and  weak turbulence energy cascades is also
phenomenologically predicted for both types of waves.
In the case of moderate ion to electron temperature ratio, a new regime of KAW turbulence at scales smaller than the electron  inertial length  is obtained, where the magnetic energy spectrum  decays like $k_\perp^{-13/3}$, thus faster	than the $k_\perp^{-11/3}$ spectrum of WW turbulence.

\end{abstract}



\section{Introduction}

Exploring the dynamics of magnetized plasmas in the range of scales
extending from the ion to the
electron Larmor radii (hereafter denoted  $\rho_i$ and $\rho_e$ respectively) is of great importance in various
contexts, including magnetic reconnection  \citep{Daughton11,Treumann13,Zweibel17} and
turbulence in space plasmas such as the solar wind
\citep{Sahraoui10,Sahraoui13,Matteini17} or the auroral regions 
\citep{Chaston08}. At scales comparable to or smaller than the electron
inertial length $d_e$, electron inertia cannot be neglected,
while electron finite Larmor radius (FLR)
corrections play a role at scales approaching  $\rho_e$.
Although a fully kinetic approach is {\it a priori} required to
describe plasma dynamics at these scales, ``reduced fluid models''
can provide an interesting insight when considering relatively
small fluctuations about a Maxwellian
equilibrium state. Up to the assumptions needed to close
the fluid hierarchy, such models can indeed be obtained using
systematic asymptotic expansions in regimes where nonlinearities
are small and characteristic length scales 
appropriately selected in order to permit a rigorous estimate of the
non-gyrotropic components of the pressure tensor. 
At scales small compared to $\rho_i$, ions are mostly static,
which leads to drastic simplifications.
Concerning the electron fluid, two regimes are classically distinguished. 
Except in the relativistic regimes of large Alfv\'en velocities \citep{Kuvshinov98}, this fluid can be viewed as incompressible in the case of  whistler waves (WWs). This assumption is in particular usual in electron magnetohydrodynamics (EMHD), a limit of a
multifluid description in which the motion of the ions is neglected and the Hall term dominates
in the Ohm's law (see e.g. \citet{Kingsep90} or \citet{Bulanov92}).
If the equilibrium plasma density is uniform, quasi-neutrality then indeed  implies 
incompressibility of the electron flow. Differently, kinetic 
Alfv\'en waves (KAWs) are compressible at the sub-ion scales, with
pressure fluctuations governed by the  perpendicular
pressure balance, a description valid as long as their frequency $\omega$
satisfies $\omega \ll k_\perp v_{th i}$ where $v_{th i}$ is the ion
 thermal velocity and $k_\perp$ the perpendicular wavenumber, leading to the electron reduced magnetohydrodynamics (ERMHD) (see \citet{Schekochihin09}
for a review). The electron gyroviscous stress tensor is negligible at
scales large compared to $\rho_e$ but, except in the case of very large values of
the ion to electron temperature ratio $\tau$, this condition strongly constrains 
the validity range when scales are also supposed
to be small compared to $\rho_i$.  This point is exemplified  in \citet{TSP16}, where
comparison with the kinetic theory shows that, in the
linear regime, the accuracy of the fluid model for KAWs with $\tau$
and the electron beta parameter 
$\beta_e$ of order unity, is limited to scales such that $k_\perp
\rho_s \simeq 15$ (here, 
$\displaystyle{\rho_s=(\frac{m_i}{2 m_e})^{1/2}\rho_e}$
is the sonic Larmor radius with $m_i$ and $m_e$ refering to the mass of the ion and of the electron respectively).
At smaller scales, non-gyrotropic pressure contributions are to be
retained. Their calculation is performed perturbatively in Appendix
\ref{append:FLR-pressure}, when the coupling to the non-gyrotropic part of the heat flux
is not retained.

In the sub-ion range, the electron inertial length
$\displaystyle{d_e=(\frac{2 m_e}{m_i\beta_e})^{1/2}\rho_s}$ plays an
important role as dispersive properties of KAWs and WWs display a
qualitative change 
across this characteristic scale, thus potentially affecting 
turbulent cascades. In order to study  scales smaller than
$d_e$, but still larger than 
$\rho_e$, the parameter $\beta_e$ should be taken small.
When assuming $\beta_e=O(m_e/m_i)$ as in \citet{ZS11}, electron
inertia should be retained, but electron FLR corrections can be
neglected, except for  $\beta_e$-independent terms involved in
gyroviscous cancellation.
For values of $\beta_e$ that, although small
compared to unity, exceed $m_e/m_i$, the electron gyroviscous force
becomes relevant, and it turns out that
its computation to order $\beta_e$ at scales comparable to $d_e$
requires the expansion to
be pushed to second order in the scale separation.

The resulting equations for the electrostatic and the parallel
magnetic potentials  
must be supplemented by conditions concerning density and
temperature fluctuations. As already mentioned, neglecting density fluctuations or
prescribing perpendicular pressure balance leads to discriminate
between WWs and KAWs. Concerning temperature fluctuations, we have
considered two different regimes. The first
one assumes isothermal electrons, which leads to 
two-field models. Beyond its simplicity, such an assumption 
appears realistic for turbulent applications when not addressing
questions related to plasma 
heating, and preferable to an adiabatic assumption \citep{TSP16}.
Temperature indeed tends to be
homogenized along the magnetic field lines which, in a turbulent regime,
are expected to be stochastic \citep{Schekochihin09}.  
A more elaborate model retains dynamical equations for the
temperatures and involves a Landau fluid closure to express the
gyrotropic heat fluxes, in terms of lower-order moments, in a way
consistent with the linear kinetic theory \citep{HP90,Snyder97}. 
As discussed by \citet{Hesse04}, retaining electron heat fluxes can be an important issue in guide-field magnetic reconnection.
The models derived in this paper could indeed be most useful to address
questions such as the role of electron pressure anisotropy, Landau damping and FLR corrections in collisionless reconnection, 
the prevalence of KAWs or WWs depending on the strength of the guiding field, as discussed by \citet{Rogers01} (see also \citet{Treumann13}).

Being three-dimensional, the reduced fluid models  are also
well adapted to study turbulence dynamics at sub-ion scales.
The regime at scales smaller than $d_e$ has been
extensively studied mostly for WWs, in the framework of EMHD
\citep{Biskamp96, Biskamp99,Galtier03,Galtier15,Lyutikov13}, of an 
incompressible bi-fluid model \citep{Andres14,Andres16a,Andres16b} 
and of extended magnetohydrodynamics (XMHD) \citep{Mil17}.
Full particle-in-cell simulations of this regime were performed 
	by \citet{Gary12} and \citet{Chang13}.
A purpose of the present paper is to  address
the KAWs dynamics at scales smaller than the electron inertial length, where new
regimes of strong and weak turbulence are phenomenologically
studied. Due to the compressibility of these waves, magnetic spectra steeper
than in the case  of WWs are obtained 
for moderate values of $\tau$, a property that could be of
interest to compare to the 
fast-decaying spectra observed in the terrestrial magnetosheath \citep{Huang14}.

The paper is organized as follows. Sections \ref{sect:Faraday},
\ref{sect:pressures}  and \ref{sec:KAW-WW} provide a derivation of
reduced models for KAWs and WWs, 
including the non-gyrotropic electron pressure force, whose 
calculation is presented in Appendix \ref{append:FLR-pressure}.
Comparisons with previous estimates are made in
Appendix \ref{append:classical}. 
In Section \ref{sec:closures}, closed systems of equations resulting 
from the assumption of isothermal electrons or from 
a Landau fluid closure are presented, both in the KAWs and WWs regimes. In Section \ref{sec:linear}, accuracy of these two closures is checked against kinetic theory 
in the linear regime.
In Section \ref{sec:turb}, the isothermal models are used as a basis for a
phenomenological theory of KAWs and WWs 
turbulent cascades, at scales either large or small compared with
the electron inertial length. The influence 
of the ion to electron temperature ratio $\tau$ on the small-scale
KAWs spectral exponent is in particular 
discussed. Section  \ref{sec:conclu} is the conclusion. 

\section{Reduced form of the Faraday equation} \label{sect:Faraday}

We consider the Faraday equation for the magnetic field ${\boldsymbol B}$
\begin{equation}
  \partial_t {\boldsymbol B}  = - c \bnabla \times {\boldsymbol E}, \label{Faraday}
\end{equation}
where the electric field ${\boldsymbol E}$ is given by the generalized Ohm's law
\begin{equation}
  {\boldsymbol E} = -\frac{1}{c} {\boldsymbol u}_e \times {\boldsymbol B} -
  \frac{1}{en} \bnabla \bcdot  {\mathsfbi P}_e
  -\frac{m_e}{e} \frac{D^{(e)}}{Dt} {\boldsymbol u}_e.  \label{Eohm}
\end{equation}
Here $c$ is the speed of light, e the electron charge and ${D^{(e)}}/{Dt}= \partial_t + {\boldsymbol u}_e \bcdot \bnabla$ holds for the material
derivative associated with the electron velocity field ${\boldsymbol u}_e$.
The electron pressure tensor ${\mathsfbi P}_e$ is given by
\begin{equation}
  {\mathsfbi P}_e = p_{\perp e} {\mathsfbi I}+
  (p_{\| e} - p_{\perp e}){\boldsymbol \tau}  + {\boldsymbol \Pi}_e,
\end{equation}
where ${\mathsfbi I}$ is the identity matrix,
$\boldsymbol \tau={\widehat{\boldsymbol b}}\otimes {\widehat{\boldsymbol b}}$ (with $\widehat{\boldsymbol b}={\boldsymbol B}/|{\boldsymbol B}|$), $p_{\| e}$ and $p_{\perp e}$ are the gyrotropic parallel and perpendicular electron pressures
and   ${\boldsymbol \Pi}_e$ refers to the FLR contribution to the electron pressure tensor. As usual, $n$ holds for the electron number density.

The equation for the parallel magnetic fluctuations $B_z$ thus reads
\begin{equation}
\frac{D^{(e)}}{Dt} B_z + (\bnabla \bcdot {\boldsymbol u}_e) B_z - ({\boldsymbol B}\bcdot  \bnabla) u_{z e} -
{\widehat{\boldsymbol z}}\bcdot \bnabla \times (\frac{c}{n e}\bnabla \bcdot {\mathsfbi P}_e)
- \frac{m_e c}{e} {\widehat{\boldsymbol z}}\bcdot \bnabla \times \frac{D^{(e)}}{Dt}{\boldsymbol u}_e =0,
\label{induc}
\end{equation}
where $\widehat{\boldsymbol z}$ denotes the unit vector along the ambient magnetic field.
The divergence of the electron velocity is given by the continuity equation
\begin{equation}
\bnabla \bcdot {\boldsymbol u}_e = -\frac{1}{n} \frac{D^{(e)}}{Dt} n. \label{eq:n}
\end{equation}  

In terms of the electrostatic potential $\varphi$ and of the magnetic potential 
${\boldsymbol A}$, defined by ${\boldsymbol B} = \bnabla \times {\boldsymbol A}$,
with the Coulomb gauge $\bnabla \bcdot {\boldsymbol A} =0$, 
Eq. (\ref{Faraday}) rewrites
\begin{equation}
{\boldsymbol E} = - \bnabla \varphi -\frac{1}{c}\partial_t {\boldsymbol A}. \label{Efarad}
\end{equation}

In addition, when concentrating on sub-ion scales where the electron velocity strongly exceeds
that of the ions, the Amp\`ere equation 
\begin{equation}
\bnabla \times {\boldsymbol B} = \frac{4\pi}{c} e n 
({\boldsymbol u}_i - {\boldsymbol u}_e),
\end{equation}
where the displacement current is neglected, reduces to
\begin{equation}
{\boldsymbol u}_e =  \frac{c}{4\upi en} \Delta {\boldsymbol A}. \label{ampere}
\end{equation}

We consider in the following fluctuations about a homogeneous
equilibrium state characterized by a number density $n_0$, isotropic
ion and electron temperatures $T_{i0}$ and $T_{e0}$ and a guide
(or ambient) field of magnitude  $B_0$ taken in the $z$-direction.
In order to deal with dimensionless quantities, we rescale time by 
the inverse ion gyrofrequency $\Omega_i = e B_0/(m_i c)$, velocities by the sound speed
$c_s = (T_{e0}/m_i)^{1/2}$, space coordinates by the sonic Larmor radius $\rho_s = c_s /\Omega_i$,
density by the equilibrium density $n_0$, magnetic field by the equilibrium field $B_0$,
electric potential by $T_{e0}/e$, parallel magnetic potential by $B_0 \rho_s$, ion pressures
by $n_0 T_{i0}$, electron pressures by $n_0 T_{e0}$, parallel and
perpendicular electron heat fluxes by $c_s n_0 T_{e0}$ and  fourth-rank moments by $n_0T_{e0}^2/m_e$. For convenience, we keep the same
notation for the rescaled fields. We also define the non-dimensional parameters
$\tau= T_{i0}/ T_{e0}$, $\delta= \sqrt{m_e/m_i}$ and $\beta_e = 8\upi n_0 T_{e0}/B_0^2$.

For a given vector 
	${\boldsymbol V}$, we denote by $V_\|$ and  
	${\boldsymbol V}_\perp$ the components 
parallel and perpendicular to the local magnetic field 
(except for the magnetic field, for which 	${\boldsymbol B}_\perp$ refers to the component perpendicular to the ambient field).
When the magnetic field
distortions are small, $V_\| = V_z$ to leading order, except when 
${\boldsymbol V}$ is quasi-perpendicular  (as it is the case for the 
electric field and also for the gradient operator). 

We assume a weakly nonlinear regime characterized by 
	two small parameters $\varepsilon$ and $\mu$
	such that $u_{\perp e}= O(\varepsilon)$ and $\bnabla_\perp =O(1/\mu)$. We furthermore assume that 
	the parallel 
	($\nabla_\| = {\widehat{\boldsymbol b}}\bcdot \bnabla$) and
	perpendicular ($\bnabla_\perp $) gradients with respect to the local magnetic field satisfy  $\nabla_\| \ll  \bnabla_\perp$,
	with scalings depending 
	on the type of wave, as specified below.  
In this asymptotics, 
	$\nabla_\|= \partial_z - [A_\|, \cdot]$, where the bracket of two scalar functions $f$ and $g$ is defined as $[f, g]= {\bhatz} \cdot (\bnabla f \times \bnabla g)$. 
Concentrating on scales around $d_e$ leads to assume
$\delta^2/(\beta_e \mu^2) =1$. At these scales, the condition (in
dimensional units)
$k_\perp\rho_e\ll 1$ reduces to $\beta_e\ll 1$.

We here consider  KAWs and WWs.
Based on their dispersion relations, we are led to prescribe that
for both kinds of waves,
$\partial_t = O(\varepsilon/\mu)$,
$\varphi   = O( \mu \varepsilon)$,
$B_z =O( \beta_e \mu \varepsilon)$.
Differently, we have  $\nabla_\| = O(\delta\varepsilon/\mu) =  O(\beta_e^{1/2}\varepsilon)$,
$A_\| = O(\delta \mu \varepsilon) = O(\beta_e^{-1/2} \delta^2\varepsilon)$
and  $u_{\|e} = O(\mu\varepsilon/\delta) =O(\beta_e^{-1/2}\varepsilon)$
for the KAWs, while for the WWs, 
$\nabla_\| = O(\beta_e\varepsilon)$, $A_\| = O(\beta_e \mu^2 \varepsilon) =  O(\delta^2\varepsilon)$
and  $u_{\|e} = O(\varepsilon)$. Furthermore, $ B_z /|{\boldsymbol
  B}_\perp| = O(\beta_e^{1/2})$ for the KAWs, but is of order unity
for the WWs (a condition which is not inconsistent with the Coulomb gauge, since  $|{\boldsymbol A}_\perp|$ and $A_\|$ can be comparable when the angle between ${\boldsymbol A}_\perp$ and ${\boldsymbol k}_\perp$ is close to
	$\pi/2$). Similarly, the density, pressure and temperature fluctuations are $O(\mu\varepsilon)$
for the KAWs, while for the WWs the density fluctuations are negligible 
and the temperature fluctuations
 are $O(\beta_e \varepsilon\mu$). For both types of waves,
the gyrotropic heat fluxes
$q_{\| e}$ and $q_{\perp e}$ scale like the parallel velocity $u_{\| e}$ and the fourth-rank cumulants
like the temperature fluctuations.

Under the above assumptions, a drift expansion of the transverse electron velocity gives
(in terms of the rescaled variables)
\begin{equation}
{\boldsymbol u}_{\perp e} = {\widehat{\boldsymbol z}}\times
(\bnabla_\perp (\varphi -p_{\perp e}) - \bnabla\bcdot {\boldsymbol \Pi}_e)\label{eq:drift1} 
\end{equation}
where, to leading order (see Appendix \ref{append:FLR-pressure}), $\bnabla\bcdot {\boldsymbol \Pi}_e= (\delta^2/2) \bnabla \omega_{ze}$,
in terms of the parallel electron vorticity $\omega_{ze}$. We write
\begin{equation}
{\boldsymbol u}_{\perp e} = {\widehat{\boldsymbol z}}\times \bnabla_\perp \varphi^*  \label{uperp}
\end{equation}
with
\begin{equation}
  \varphi^* = \varphi -p_{\perp e} - \frac{\delta^2}{2} \omega_{ze}. \label{eq:phi*}
\end{equation}
In Eq. (\ref{eq:phi*}) and hereafter, the various fields refer to fluctuations, except for the pressure in  primitive equations.

Furthermore, using  (\ref{ampere}),
\begin{eqnarray}
  && u_{\|e} =  \frac{2}{\beta_e}\Delta_\perp A_\|  \label{uparApar} \\
  &&\omega_{ze} = \Delta_\perp \varphi^*\\
  &&\varphi^*= \frac{2}{\beta_e}  B_z \label{phi-Bz} \\
  && (1 + \frac{\delta^2}{2} \Delta_\perp) \varphi^* = \varphi - p_{\perp e}. \label{phi-phi*}
\end{eqnarray}
It follows that
\begin{equation}
\frac{D^{(e)}}{Dt} = \frac{d}{dt} -[p_{\perp e}+ \frac{\delta^2}{\beta_e} \Delta_\perp B_z, \bcdot],
\end{equation}
where $\displaystyle{\frac{d}{dt}=\partial_t +[\varphi,\bcdot]}$.

The projection on the parallel direction of Eq. (\ref{Efarad}), 
\begin{equation}
	\partial_t A_\| + c \nabla_\| \varphi = - c E_\|, \label{A1}
\end{equation}
rewrites 
\begin{equation}
\frac{d}{dt} \left (1 -\frac{2\delta^2}{\beta_e}\Delta_\perp\right )A_\| + \partial_z \varphi - \nabla_\| p_{\|e}
+\delta^2 [p_{\perp e}+\frac{\delta^2}{2}\omega_{ze}, u_{\|e}] - {\widehat{\boldsymbol b}}\bcdot (\bnabla \bcdot  {\boldsymbol \Pi}_e)=0, \label{Aeq}
\end{equation}
where the FLR contributions ${\widehat{\boldsymbol b}}\bcdot (\bnabla \bcdot  {\boldsymbol \Pi}_e)
= \bhatb\bcdot \bnabla\bcdot {\boldsymbol \Pi}^{(1)}_e + \bhatb\bcdot \bnabla\bcdot {\boldsymbol \Pi}^{(2)}_e$
are given by Eqs.  (\ref{eq:divPI_par}), (\ref{eq:phis-expression}),
(\ref{bdivPi2}) and (\ref{eq:divPI_par2}).
Equation (\ref{Aeq}) rewrites 
\begin{eqnarray}
&&\frac{d}{dt} \left (1 -\frac{2\delta^2}{\beta_e}\Delta_\perp+\frac{2\delta^4}{\beta_e}\Delta^2_\perp\right )A_\| + \partial_z \varphi - \nabla_\| p_{\|e}
-[p_{\perp e},\frac{2\delta^4}{\beta_e}\Delta^2_\perp A_\|]\nonumber\\
&&  \qquad +\delta^2[B_z,u_{\| e}+q_{\perp e}] 
 -\delta^2 \nabla_\| \Delta_\perp \varphi^*  - \delta^2 \sum_{i=x,y} [\partial_i\varphi^*,(1-\frac{2\delta^2}{\beta_e}\Delta_\perp)\partial_i A_\|]\nonumber\\
&&  \qquad +\delta^4 \frac{d}{dt}\Delta_\perp q_{\perp
   e}-\delta^4[p_{\perp e}, \Delta_\perp q_{\perp e}]
 -\delta^4[\Delta_\perp\varphi^*,q_{\perp  e}]=0, \label{Aeq2}
\end{eqnarray}
where the gyroviscous cancellation eliminates the $\delta^2 [p_{\perp e}+\frac{\delta^2}{2}\omega_{ze} ,  u_{\|e}]$ term.

We now turn to Eq. (\ref{induc}) where, assuming an isotropic equilibrium state, we write
\begin{equation}
\bnabla \bcdot {\mathsfbi P}_e = \nabla_\perp p_{\perp e} + \nabla_\| p_{\| e} \bhatb + \bnabla\bcdot{\boldsymbol \Pi}_e.
\end{equation}
At the order of the present expansion, we have
\begin{equation}
  {\widehat{\boldsymbol z}}\bcdot \bnabla \times \frac{D^{(e)}}{Dt}{\boldsymbol u}_e = \frac{D^{(e)}}{Dt} \omega_{ze}=
  \frac{2}{\beta_e} \frac{D^{(e)}}{Dt}\Delta_\perp B_z \label{eq:curlaccel}
\end{equation}
and
\begin{eqnarray}
&&{\widehat{\boldsymbol z}}\bcdot \bnabla \times (\bnabla \bcdot
{\boldsymbol \Pi}_e)= \frac{2\delta^2}{\beta_e} [p_{\perp e}-B_z,\Delta_\perp B_z]+
\delta^2\sum_{i=x,y} [\partial_i (p_{\perp
    e}-B_z),\partial_i\varphi^*]\nonumber \\
&& \qquad -\frac{\delta^2}{2}\Delta_\perp  \bnabla\bcdot
  \boldsymbol{u}_e -\frac{\delta^2}{2}\nabla_\| \Delta_\perp u_{\|
    e}  -\delta^2\nabla_\| \Delta_\perp q_{\perp e}
+ \delta^2\sum_{i=x,y} [\partial_i A_\|, \partial_i q_{\perp e}].
\end{eqnarray}
Using  Eq. (\ref{DnxDPi1}), we obtain
\begin{eqnarray}
  &&  \frac{d}{dt} \Big ( (1 -
  \frac{2\delta^2}{\beta_e}\Delta_\perp) B_z\Big ) - [p_{\perp e} + 
    \frac{\delta^2}{\beta_e} \Delta_\perp B_z, B_z+n] 
  + ( 1+\frac{\delta^2}{2} \Delta_\perp ) (\bnabla \bcdot {\boldsymbol u}_e)\nonumber \\
  && -\nabla_\| \Big (1 -\frac{\delta^2}{2} \Delta_\perp\Big ) u_{\| e} 
  +\delta^2 [B_z, \Delta_\perp \varphi^*] - \delta^2 \sum_{i=x,y}
  [\partial_i p_{\perp e}, \partial_i \varphi^*] \nonumber \\
  &&+\delta^2\nabla_\| \Delta_\perp q_{\perp e}
- \delta^2\sum_{i=x,y} [\partial_i A_\|, \partial_i q_{\perp e}]   =0
  \label{eq:Bz}
\end{eqnarray}
or, when using Eq. (\ref{eq:n}),
\begin{eqnarray}
  &&  \frac{d}{dt} \Big ( (1 - \frac{2\delta^2}{\beta_e}\Delta_\perp) B_z -(1+\frac{\delta^2}{2}\Delta_\perp) n \Big)
  - [p_{\perp e} + \frac{3\delta^2}{\beta_e} \Delta_\perp B_z, B_z]-\frac{\delta^2}{2} [\Delta_\perp \varphi, n] 
    \nonumber \\
  && +\frac{\delta^2}{2} \Delta_\perp[p_{\perp e} + \frac{\delta^2}{\beta_e}
    \Delta_\perp B_z, n] 
  - \frac{2}{\beta_e}\nabla_\| (1 -\frac{\delta^2}{2} \Delta_\perp )  \Delta_\perp A_\|
  +\delta^2\sum_{i=x,y}[\partial_i n,\partial_i \varphi]\nonumber \\
  &&  - \delta^2[\partial_i p_{\perp e}, \partial_i\varphi^*]
  +\delta^2\nabla_\| \Delta_\perp q_{\perp e}
- \delta^2\sum_{i=x,y} [\partial_i A_\|, \partial_i q_{\perp e}]   =0. \label{nBzeq2}
\end{eqnarray}

\section{Pressure equations}\label{sect:pressures}

The equations for the electron gyrotropic pressures read
\begin{eqnarray}
&&\frac{D^{(e)}}{Dt} p_{\|e} +  p_{\|e}\bnabla \bcdot {\boldsymbol u}_e + 2  p_{\|e}
\nabla_\| {\boldsymbol u}_e\bcdot \bhatb + \bnabla \bcdot {\mathsfbi q}_e \, {\boldsymbol :} \, {\boldsymbol \tau}  
+\Big [({\boldsymbol \Pi}_e \bcdot \bnabla {\boldsymbol u_e})^S \, {\boldsymbol :} \, {\boldsymbol \tau}  -
{\boldsymbol \Pi}_e  \, {\boldsymbol :} \,\frac{d {\boldsymbol \tau} }{dt}\Big ] =0 \label{pparal} \\
&&\frac{D^{(e)}}{Dt} p_{\perp e } +  2p_{\perp e }\bnabla\bcdot {\boldsymbol u}_e  -
p_{\perp e} \nabla_\| {\boldsymbol u}_e\bcdot \bhatb +
\frac{1}{2} \bnabla \bcdot {\mathsfbi q}_e  \, {\boldsymbol :} \, {\mathsfbi n} +\frac{1}{2}\Big [
({\boldsymbol \Pi}_e \bcdot \bnabla {\boldsymbol u_e})^S  \, {\boldsymbol :} \, {\mathsfbi n}  +
{\boldsymbol \Pi}_e  \, {\boldsymbol :} \,\frac{d {\boldsymbol \tau} }{dt} \Big ] =0,\nonumber \\
&&\label{pperp}
\end{eqnarray}
where ${\mathbf n}= {\mathbf I} -{\boldsymbol \tau}$ and the electron heat flux tensor  ${\mathsfbi q}_e$
can be  written 
${\mathsfbi q}_e= {\mathsfbi S}_e + {\boldsymbol \sigma}_e$. Here the  tensor 
${\boldsymbol \sigma}_e$ obeys the conditions
${\boldsymbol \sigma}_e  \, {\boldsymbol :} \,{\mathsfbi n} =0$ and 
${\boldsymbol \sigma}_e  \, {\boldsymbol :} \,{\boldsymbol \tau}=0$.
The elements of the tensor ${\mathsfbi S}_e$ are classically expressed (see e.g.  \cite{GPS05})
in terms of the components of
two vectors ${\boldsymbol S}^\|_e$ and ${\boldsymbol S}^\perp_e$ defined by 
${\boldsymbol S}^\|_e = {\mathsfbi q}_e  \, {\boldsymbol :} \, {\boldsymbol \tau}$ and 
${\boldsymbol S}^\perp_e =  (1/2)\ {\mathsfbi q}_e  \, {\boldsymbol :} \, {\mathsfbi n}$
that measure the directional fluxes of the parallel and perpendicular heats respectively. 
The usual perpendicular and parallel gyrotropic heat fluxes are given by  
$q_{\perp e}= {\boldsymbol S}_e^\perp \bcdot {\boldsymbol b}$ and $q_{\| e}= 
{\boldsymbol S}_e^\| \bcdot {\boldsymbol b}$, and thus correspond to the fluxes along the magnetic field.
We  write ${\boldsymbol S}_r^\perp= q_{\perp e} {\boldsymbol b} + {\boldsymbol S}^\perp_{ \perp e}$ and 
${\boldsymbol S}_e^\|= q_{\| e} {\boldsymbol b} + {\boldsymbol S}^\|_{\perp e}$.

In the present asymptotics, the contribution to the pressure equations of the
tensor ${\boldsymbol \sigma}_e$, whose expression, given in \cite{Ramos05a}
and also in \cite{SP15}, is negligible.
To leading order, we are thus led to  write 
\begin{eqnarray}
&&(\nabla \bcdot {\mathbf q}_e )  \, {\boldsymbol :} \, {\boldsymbol \tau} = 
\nabla_\| q_{\| e}  + \nabla \cdot {\mathbf S}^\|_{\perp e} \\
&& \frac{1}{2} (\nabla \bcdot {\mathbf q}_e)  \, {\boldsymbol :} \, {\mathbf n}   =
\nabla_\| q_{\perp e}   + \nabla \cdot {\mathbf S}_{\perp e}^\perp.   
\end{eqnarray}
Here, the non-gyrotropic heat fluxes contributions, described by the vectors ${\mathbf S}^\|_{\perp e}$
and ${\mathbf S}_{\perp e}^\perp$, are obtained by changing sign in the equations 
Eqs. (3.6) and (3.7) of \cite{SP15} for the ion heat flux vectors. We obtain, at the order of the asymptotics, 
\begin{eqnarray}
  &&{\mathbf S}_{\perp e}^\| = - \frac{1}{B} {\widehat {\mathbf b}}
  \times (p_{\perp e}\bnabla T_{\| e} +  2 \delta^2 q_{\perp e} \bhatb \btimes {\boldsymbol \omega}_e +
  \bnabla {\widetilde r}_{\| \perp e}) \\
  &&{\mathbf S}_{\perp e}^\perp = - \frac{2}{B} {\widehat {\mathbf b}}
  \times (p_{\perp_e} \bnabla T_{\perp e} + \bnabla {\widetilde r}_{\perp \perp e}),
\end{eqnarray}
where the ${\widetilde r}$ functions refer to the gyrotropic components of the fourth-rank cumulants.

Among the terms associated with the work of the non-gyrotropic pressure force in (\ref{pparal}) and (\ref{pperp}), the one involving
the time derivative scales like $O(\varepsilon^3)$ and is thus negligible. 
In the other terms, only the linear part of ${\boldsymbol \Pi}_e$ is possibly relevant at the
prescribed order, but it is easily checked that the corresponding contribution
in fact vanishes. 

Using the electron continuity equation to eliminate the velocity divergence, we are led to write
the equations for the temperature fluctuations
\begin{align}
&\frac{d}{dt} T_{\|e} - [B_z, T_{\| e} + {\widetilde r}_{\|\perp e}] + \nabla_\| (2 u_{\| e} + q_{\| e})
  -\frac{\delta^2}{\beta_e} [\Delta_\perp B_z, T_{\| e} ] + 2 \delta^2 [q_{\perp e } , u_{\| e}] = 0 \label{Tpar} \\
&\frac{d}{dt} (T_{\perp e} - n) - 2 [B_z, T_{\perp e} + {\widetilde r}_{\perp\perp e}] +
  \nabla_\| (q_{\perp e}-u_{\| e}) -\frac{\delta^2}{\beta_e} [\Delta_\perp B_z, T_{\perp e} -n]
  =0, \label{Tperp}
\end{align}
where $u_{\|e}$ and $B_z$ are given by Eqs. (\ref{uparApar}) and (\ref{phi-Bz}) respectively.

\section{KAWs or WWs, depending on the compressibility level} \label{sec:KAW-WW}

At this stage, it is important to note that, beyond the estimate of 
the gyrotropic electron pressure fluctuations which are prescribed by the closure
assumption for the fluid hierarchy, the system given by Eqs. (\ref{Aeq2}) and (\ref{nBzeq2})
involves three unknown quantities $A_\|$,
$\varphi$ and $n$. An additional relation is obtained when specifying
the type of waves the system is describing. KAWs and WWs are indeed 
characterized by different levels of compressibility.

In the case of KAWs,
the plasma is in pressure balance in the transverse direction. 
This condition, which consists in neglecting the inertial term compared to the pressure gradient in the equation for the transverse total momentum reads (for typical ion transverse
velocity $\delta v_{\perp i}$ and density fluctuations $\delta n$), 
\begin{equation}
\frac{\omega}{k_\perp} \frac{\delta v_{\perp i}}{v_{th i}} \ll
v_{th i} \frac{\delta n}{n}.
\end{equation}
In the above estimate, in spite of the fact that 
$|{\boldsymbol u}_i| \ll |{\boldsymbol u}_e|$, $m_i \delta v_{\perp i}$
dominates $m_e \delta v_{\perp e}$, at least close to the ion scale
where ${\delta v_{\perp i}}/{v_{th i}}$ and ${\delta n}/{n}$
are comparable, and where one recovers the classical condition $\omega\ll
k_\perp v_{th i}$ for the existence of KAWs. At smaller scales,
where the ion velocity fluctuations decrease due to ion
 demagnetization, this condition 
becomes in fact less stringent.
The pressure balance then reads
\begin{equation}
(\bnabla\bcdot (\tau {\mathsfbi P}_i + {\mathsfbi P}_e))_\perp -
\frac{2}{\beta_e} (\bnabla \times {\boldsymbol B}) \times {\widehat{\boldsymbol z}} =0, \label{FarOhm}
\end{equation}
where the index $i$ refers to the ions.  Applying the transverse divergence, we find
\begin{equation}
\Delta_\perp (\tau p_{\perp i} + p_{\perp e} ) + \bnabla_\perp \bcdot (\bnabla \bcdot (\tau{\boldsymbol \Pi}_i  +{\boldsymbol \Pi}_e))
+\frac{2}{\beta_e}\Delta_\perp B_z =0. \label{balance}
\end{equation}
As shown in \cite{TSP16}, at the sub-ion scales, ${\boldsymbol \Pi}_i$ is negligible and the ions are isothermal
(making in the present units electron pressure and density fluctuations equal).
To leading order in $\varepsilon$, Eq. (\ref{balance}) rewrites
\begin{equation}
{\tau n + p_{\perp e} + \frac{\delta^2}{2} \Delta \varphi^* + \frac{2}{\beta_e} B_z}=0.
\end{equation}
Using Eqs. (\ref{phi-Bz}) and (\ref{phi-phi*}), it is then easily checked that
\begin{equation}
n = -\frac{1}{\tau} \varphi.
\end{equation}
It results that
\begin{eqnarray}
&&n=-\frac{1}{1+\tau}\left ( T_{\perp e} + (1+\frac{\delta^2}{2}\Delta_\perp) \varphi^* \right ) \label{eq:nphi*}\\
&&\varphi=\frac{\tau}{1+\tau}\left (T_{\perp e} + (1+\frac{\delta^2}{2}\Delta_\perp) \varphi^* \right ).\label{eq:phiphi*}
\end{eqnarray}

A different behavior of the density holds in the case of the WWs. The linear kinetic theory
(see e.g. \citet{Gary09}) shows that the electron compressibility of
WWs is significantly smaller
than that of the KAWs, at least for $\beta_e$ 
small enough compared to unity and  scales small compared to $d_i$.
 This suggests that we may
assume $\bnabla \bcdot {\boldsymbol u}_e$ smaller than $\varepsilon^2$, and thus may neglect the
density fluctuations. This regime formally corresponds to taking the limit $\tau\to\infty$ in Eqs. (\ref{eq:nphi*})-(\ref{eq:phiphi*}). 
Note however than in auroral zones where the Alfv\'en velocity can exceed several
thousands of km/s, quasi-neutrality no longer holds at large enough wavenumber, making electron
density fluctuations relevant for WW dynamics \citep{Kuvshinov98}.

In the next Section, two different closures of the fluid hierarchy will be examined,
namely the isothermal case, a regime often considered in turbulence studies, and a Landau fluid closure.

\section{Closure assumptions}\label{sec:closures}

\subsection{General form of the four-field model}

Two asymptotics which are similar but not identical can be distinguished.
When considering scales comparable to $d_e$, we should take $\beta_e$ as the expansion parameter.
We refer in the following to this regime as the small $\beta_e$ regime.
Differently, when considering $\beta_e$ of order unity, a gradient expansion is performed in terms of $\delta^2\Delta_\perp$
(hereafter referred to as large-scale regime).
In fact both approaches can be captured simultaneously, but this leads us to retain subdominant
terms of order $\beta_e \delta^2\Delta_\perp = O(\beta_e^2)$ in the former regime and
of order $\delta^4 \Delta_\perp^2/\beta_e$ in the latter.

In the KAW regime, using Eqs. (\ref{eq:nphi*})-(\ref{eq:phiphi*}),  Eqs. (\ref{Aeq2}) and (\ref{nBzeq2}) rewrite,
\begin{eqnarray}
&&\partial_t \left ( (1 -\frac{2\delta^2}{\beta_e}\Delta_\perp+\frac{2\delta^4}{\beta_e}\Delta^2_\perp )A_\|
+\delta^4 \Delta_\perp q_{\perp e}\right )
+\nabla_\|(T_{\perp e} - T_{\|e}) + \nabla_\|(1- \frac{\delta^2}{2}\Delta_\perp)\varphi^*\nonumber\\
&& + [\varphi^*, \delta^2\Delta_\perp A_\|+\frac{2\delta^4}{\beta_e}\Delta_\perp^2 A_\|]
- \frac{\tau}{1+\tau}[T_{\perp e} + (1+ \frac{\delta^2}{2}\Delta_\perp)\varphi^*,\frac{2\delta^2}{\beta_e}\Delta_\perp A_\|]
\nonumber \\
&& - \delta^2 \sum_{i=x,y} [\partial_i\varphi^*,(1-\frac{2\delta^2}{\beta_e}\Delta_\perp)\partial_i A_\|]
+ [\varphi^*, \frac{\beta_e}{2}\delta^2 q_{\perp e}+\delta^4\Delta_\perp q_{\perp e}] -\delta^4[\Delta_\perp\varphi^*,q_{\perp  e}]=0,\nonumber\\
&&  \label{eq:A-unif}
\end{eqnarray}
and
\begin{eqnarray}
&&  \partial_t \Big ( \frac{\beta_e}{2}(1 - \frac{2\delta^2}{\beta_e}\Delta_\perp) \varphi^*
+\frac{1}{1+\tau}(1+\delta^2\Delta_\perp) \varphi^*
+\frac{1}{1+\tau}(1+\frac{\delta^2}{2}\Delta_\perp) T_{\perp e} \Big) \nonumber \\
&& - \frac{2}{\beta_e}\nabla_\| (1 -\frac{\delta^2}{2} \Delta_\perp )  \Delta_\perp A_\|
+\delta^2\nabla_\| \Delta_\perp q_{\perp e} +\frac{2\tau+1}{2(\tau+1)} [\delta^2\Delta_\perp \varphi^*, T_{\perp e}] 
+\delta^2\sum_{i=x,y}[\partial_i \varphi^*, \partial_i T_{\perp e}] \nonumber \\
&&   -[\varphi^*, (\frac{\tau}{1+\tau} -\frac{\beta_e}{2})\delta^2 \Delta_\perp\varphi^*
- \frac{1}{1+\tau} \frac{\delta^2}{2}\Delta_\perp T_{\perp e}]
- \delta^2\sum_{i=x,y} [\partial_i A_\|, \partial_i q_{\perp e}]   =0.\label{eq:phi*-unif}
\end{eqnarray}
Note that retaining $O(\beta_e)$ contributions make temperature anisotropy relevant
in Eqs. (\ref{eq:A-unif}) and (\ref{eq:phi*-unif}), while neglecting these terms and using $\varphi$ instead of $\varphi^*$ eliminates $T_{\perp e}$ from these equations.
	
Equations (\ref{Tpar}) and (\ref{Tperp}) become
\begin{eqnarray}
&&\partial_t T_{\|e}  + \nabla_\| (\frac{4}{\beta_e}\Delta_\perp A_\| + q_{\| e}) + \frac{\tau}{1+\tau}[T_{\perp e} + (1+\frac{\delta^2}{2}\Delta_\perp)\varphi^*, T_{\| e}] \nonumber \\
&&   - \frac{\beta_e}{2}[\varphi^*, T_{\| e} + {\widetilde r}_{\|\perp e}]-[\frac{\delta^2}{2}\Delta_\perp\varphi^*,T_{\| e}]  + \frac{4 \delta^2}{\beta_e} [q_{\perp e } , \Delta_\perp A_\|] = 0  \label{eq:Tparal2}\\
&&\partial_t \left (\frac{2+\tau}{1+\tau}T_{\perp e} + \frac{1}{1+\tau} (1+\frac{\delta^2}{2}\Delta_\perp)\varphi^*\right)
+\nabla_\| (q_{\perp e}-\frac{2}{\beta_e}\Delta_\perp A_\|)\nonumber \\
&& - \frac{\tau}{1+\tau} [T_{\perp e}, (1+\frac{\delta^2}{2}\Delta_\perp) \varphi^*] - \beta_e [\varphi^*, T_{\perp e}
  + {\widetilde r}_{\perp\perp e}]-\frac{\delta^2}{2} [\Delta_\perp \varphi^*,\frac{2+\tau}{1+\tau}T_{\perp e}
+ \frac{1}{1+\tau} \varphi^*]=0 .\nonumber \\
&&  \label{eq:Tperp2}
\end{eqnarray}
In the limit $\delta=0$, Eqs. (\ref{eq:A-unif})-(\ref{eq:Tperp2}) reduce to Eqs. (3.72)-(3.78) of \citet{TSP16}.
In the adiabatic limit (where $q_{\| e}$, $q_{\perp e}$, $ {\widetilde r}_{\|\perp e}$ and ${\widetilde r}_{\perp\perp e}$
are taken equal to zero), this system possesses a conserved energy, a property easily established when noting
that the brackets involving gradients are eliminated within the integrals by using equalities of the type
\begin{eqnarray}
&& 2\int \Delta_\perp A_\| \sum_{i=x,y}[\partial_i\varphi^*, \partial_i A_\|] d {\boldsymbol x} =
\int \left (\Delta_\perp^2 A_\| [\varphi^*, A_\|]
-\Delta_\perp A_\| [\Delta_\perp \varphi^*, A_\|] \right) d {\boldsymbol x}\\
&& \int \Delta_\perp A_\| \sum_{i=x,y}[ \partial_i \varphi^*, \partial_i \Delta_\perp A]  d {\boldsymbol x} =
\int \Delta_\perp^2 A [\varphi^*, \Delta_\perp A_\|] d {\boldsymbol x}.
\end{eqnarray}
These identities are obtained by expanding $\Delta_\perp [\varphi^*, f ]$ within the equality
\begin{equation}
\int \Delta_\perp A_\| \Delta_\perp[\varphi^*, f] d {\boldsymbol x} =
\int  \Delta_\perp^2 A_\|[\varphi^*, f] d {\boldsymbol x},
\end{equation}
where $f$ holds for $A_\|$ or $\Delta_\perp A_\|$,
and using the identity $\int f [g, h] d {\boldsymbol x} =  \int h [f, g]d {\boldsymbol x}$.
The energy reads
\begin{eqnarray}
&&{\mathcal E}_{KAW}=\frac{1}{2} \int \frac{2}{\beta_e}\Big (
|\bnabla A_\| |^2 + \frac{2\delta^2}{\beta_e} (\Delta_\perp A_\|)^2
+ \frac{2\delta^4}{\beta_e} |\bnabla\Delta_\perp A_\||^2 \Big )+\Big (\frac{1}{\tau+ 1} + \frac{\beta_e}{2}\Big ) \varphi^{*2} \nonumber\\
&& +\frac{\tau}{1+\tau} \delta^2 |\bnabla \varphi^*|^2 + \frac{T_{\|e}^2}{2}+\frac{2+\tau}{1+\tau}{T_{\perp e}^2}+\frac{2}{1+\tau}T_{\perp e}
(1+\frac{\delta^2}{2}\Delta_\perp) \varphi^*  d {\boldsymbol x}.\label{eq:energy-LF}
\end{eqnarray}

Writing that, to leading order, $ \varphi^*=-(T_{\perp e}+(1+\tau)n) $, it can be shown that the internal energy constributions in
Eq. (\ref{eq:energy-LF}) coincide with those of Eq. (3.39) of \citet*{TSP16}, once the various
fields are transformed from the gyrofluid to the particle formulation.

The system of equations describing the WWs dynamics is conveniently obtained by taking the limit $\tau\to \infty$.  In this regime, the temperature fluctuations are subdominant by a factor $\beta_e$ in the resulting system.

\subsection{The isothermal case} \label{sec:isothermal}

Isothermal electrons  is a good approximation when one does not 
focus on dissipative effects, as long as $k_\|\ll k_\perp$ and $k_\perp \rho_e\ll 1$, as discussed in
\citet{Schekochihin09}.

\subsubsection{KAWs reduced model}

Taking  temperature fluctuations and heat fluxes equal to zero, one obtains, in the small $\beta_e$ regime,
\begin{eqnarray}
  &&\partial_t \left (1-{\frac{2\delta^2}{\beta_e} \Delta_\perp} +{\frac{2\delta^4}{\beta_e}\Delta^2_\perp }\right )A_\|
+ \nabla_\|(1-{\frac{\delta^2}{2} \Delta_\perp})\varphi^*
+{[\varphi^*, \delta^2\Delta_\perp A_\|+\frac{2\delta^4}{\beta_e}\Delta^2_\perp A_\|]} \nonumber \\
  &&  - \frac{\tau}{\tau+1} [(1+ {\frac{\delta^2}{2}\Delta_\perp}) \varphi^*, {\frac{2\delta^2}{\beta_e} \Delta_\perp A_\|} ]
  -{\delta^2 \sum_{i=x,y} [\partial_i \varphi^*,(1-\frac{2\delta^2}{\beta_e}\Delta_\perp)\partial_i A_\|]}=0   \label{eq:A-KAW-iso2}\\
  &&\partial_t \left (\frac{\beta_e}{2}(1-{\frac{2\delta^2}{\beta_e}\Delta_\perp)} +
  \frac{1}{\tau + 1}(1+{\delta^2 \Delta_\perp} )\right )\varphi^* -\frac{2}{\beta_e} \nabla_\|(1-{\frac{\delta^2}{2}\Delta_\perp} ) \Delta_\perp A_\|\nonumber \\
  && -{\frac{\tau}{1 + \tau}[\varphi^*, \delta^2\Delta_\perp \varphi^*]}  =0, \label{eq:phi-KAW-iso2}
\end{eqnarray}

while at finite $\beta_e$ and scales large compared to $\rho_e$, the system reduces to
\begin{eqnarray}
&&\partial_t \left (1-\frac{2\delta^2}{\beta_e} \Delta_\perp \right )A_\|
+ \nabla_\|(1-\frac{\delta^2}{2} \Delta_\perp)\varphi^*
+ [\varphi^*, \delta^2\Delta_\perp A_\|] \nonumber \\
&&  - \frac{\tau}{\tau+1} [\varphi^*, \frac{2\delta^2}{\beta_e} \Delta_\perp A_\| ]
-\delta^2 \sum_{i=x,y} [\partial_i \varphi^*,\partial_i A_\|]=0   \\
&&\partial_t \left (\frac{\beta_e}{2}(1-\frac{2\delta^2}{\beta_e}\Delta_\perp) +
\frac{1}{\tau + 1}(1+ \delta^2 \Delta_\perp )\right )\varphi^* -\frac{2}{\beta_e} \nabla_\|(1-\frac{\delta^2}{2}\Delta_\perp ) \Delta_\perp A_\| \nonumber \\
&& - (\frac{\tau}{1+\tau}-\frac{\beta_e}{2})[\varphi^*, \delta^2\Delta_\perp \varphi^*]=0. 
\end{eqnarray}

In the small $\beta_e$ regime, the energy, which reduces to 
\begin{eqnarray}
 {\mathcal E}_{KAW}&=&\frac{1}{2} \int \Big \{ \Big (\frac{1}{\tau+ 1} + \frac{\beta_e}{2}\Big ) \varphi^{*2} +
  \frac{\tau}{1+\tau} \delta^2 |\bnabla \varphi^*|^2 + \frac{2}{\beta_e}\Big (
  |\bnabla A_\| |^2  \nonumber \\
&& + \frac{2\delta^2}{\beta_e} (\Delta_\perp A_\|)^2
  + \frac{2\delta^4}{\beta_e} |\bnabla\Delta_\perp A|^2 \Big ) \Big \} 
  d {\boldsymbol x},
\end{eqnarray}
is conserved during the time evolution. In the case of finite $\beta_e$, the term 
proportional to $\delta^4$ is absent in the definition of the energy.

In the limit $\beta_e \to 0$, $\tau \to \infty$ 
	with $\tau = O(\beta_e^{-1})$, the system considered to leading order at scales
	 comparable to the electron inertial length reduces to Eqs. (19)-(20) of 
	 \citet{Chen-Boldyrev17}.
    
\subsubsection{WWs reduced model}

 Proceeding as above, we get in the case of small $\beta_e$ 
\begin{eqnarray}
  &&\partial_t \left (1-\frac{2\delta^2}{\beta_e} \Delta_\perp +\frac{2\delta^4}{\beta_e}\Delta^2_\perp \right )A_\|
+ \nabla_\|(1-\frac{\delta^2}{2} \Delta_\perp)\varphi^*
+ [\varphi^*, \delta^2\Delta_\perp A_\|+\frac{2\delta^4}{\beta_e}\Delta^2_\perp A_\|] \nonumber \\
  &&  -  [(1+ \frac{\delta^2}{2}\Delta_\perp) \varphi^*, \frac{2\delta^2}{\beta_e} \Delta_\perp A_\| ]
  -\delta^2 \sum_{i=x,y} [\partial_i \varphi^*,(1-\frac{2\delta^2}{\beta_e}\Delta_\perp)\partial_i A_\|]=0   \label{eq:A-WW-iso2}\\
  &&\partial_t \left (1-\frac{2\delta^2}{\beta_e}\Delta_\perp  \right )\varphi^*
  -\frac{4}{\beta_e^2} \nabla_\|(1-\frac{\delta^2}{2}\Delta_\perp ) \Delta_\perp A_\|
  + \Big (1 -\frac{2}{\beta_e}\Big )[\varphi^*, \delta^2\Delta_\perp \varphi^*]
  =0. \label{eq:phi-WW-iso2}
\end{eqnarray}

The corresponding energy reads
\begin{equation}
  {\mathcal E}_{WW}=\frac{1}{2} \int \Big \{ \varphi^{*2} +
  \frac{2\delta^2}{\beta_e} |\bnabla \varphi^*|^2 + \frac{4}{\beta_e^2}\Big (
  |\bnabla A_\| |^2 + \frac{2\delta^2}{\beta_e} (\Delta_\perp A_\|)^2
  + \frac{2\delta^4}{\beta_e} |\bnabla\Delta_\perp A|^2 \Big ) \Big \} d {\boldsymbol x}.
\end{equation}

In the finite $\beta_e$ regime, the only change in the system of equations and in the corresponding
energy consists in the absence of the $\delta^4$ contributions.

Note that in the range $\rho_i^{-1} \ll k_\perp \ll d_e^{-1}$, where both electron inertia and FLR corrections
are negligible, the equations for KAWs and for WWs formally identify, although $\varphi^*$
refers in fact to different quantities, being proportional to the electron density
in the former case and to the parallel magnetic field in the latter \citep{BHXP13}.
The resulting model is usually referred to as ERMHD (see e.g. \citet{Schekochihin09}), 
and can be viewed as the 
generalization of EMHD for low-frequency anisotropic fluctuations 
without the assumption of incompressibility.
When electron inertia is retained, but not the FLR corrections, 
the equations for the WWs appear as an  extension to the anisotropic three-dimensional regime of the 2.5D equations given in \citet{Biskamp96, Biskamp99}.

\subsection{Landau fluid closure}\label{sec:landau}

In Eqs. ({\ref{eq:A-unif})-(\ref{eq:Tperp2}) for KAWs, which are valid both in the small $\beta_e$ and the large-scale regimes (or in the equations for WWs which, as previously mentioned, are obtained by taking the limit $\tau \to \infty$),  the heat fluxes and fourth-rank cumulants are still to be specified.
A  simple closure, aimed at capturing Landau damping, in a way consistent with linear kinetic theory, is provided by directly expressing the heat fluxes in terms of lower-order fluctuations, as done in equations (3.17)-(3.19)
of \citet{SP15}, which here reduce to 
\begin{eqnarray}
&& q_{\| e}=-2\alpha H T_{\| e} \label{eq:qparal}\\
&& q_{\perp e}=-\alpha H T_{\perp e} -\alpha\delta^2 H \omega_{ze} \label{eq:qperp}
\end{eqnarray}
where, at the considered scales, the vorticity term arising in
Eq. (3.18) of the above reference has, in fact, also to be retained. Here,
$\displaystyle{\alpha=(\frac{2}{\pi})^{1/2} \delta^{-1}}$ and 
the operator $H$ denotes the negative Hilbert transform along the
magnetic field lines (see \citet{SP15} for a discussion on its
modeling).
The fourth-order cumulants are given by equations (3.21) and (3.27) of \citet{SP15}, which here
take the form
\begin{eqnarray}
&& {\widetilde r}_{\| \perp e}=-T_{\perp e}-\alpha\delta^2 H \omega_{ze} 
\label{eq:rparal}\\
&& {\widetilde r}_{\perp \perp e}=0. \label{eq:rperp}
\end{eqnarray}

In contrast with the small-scale model
discussed in \citet{TSP16}, the Reduced Landau fluid (RLF) model discussed here includes an
explicit closure of the fluid hierarchy and retains electron inertia
together with FLR corrections.

\medskip
\noindent
\textit{Remark:} The present RLF model can be extended to larger
scales by relating $\varphi$ and $n$, as in \citet{ZS11}, by means of the
more general relation provided by the
gyrokinetic Poisson equation \citep{Krommes02}
\begin{equation}
n=-\left (\frac{1-\Gamma_0(\tau   k^2_\perp)}{\tau}\right )\varphi.
\label{eq:gamma0}
\end{equation}
Here $\displaystyle{\Gamma_0(x)=e^{-x}I_0(x)}$, where $I_0$ denotes the
modified Bessel function of order $0$. In this approach, the ion response is  taken into account and the
agreement with kinetic theory improved at values of $k_\perp$ close
to unity, as discussed in Section \ref{sec:linear}.
The model then appears as a Landau fluid
version of the kinetic model of \citet{ZS11}, where FLR corrections
have been supplemented. 

\section{Linear theory} \label{sec:linear}

\begin{figure}
	\centerline{
		\includegraphics[width=0.48\textwidth]{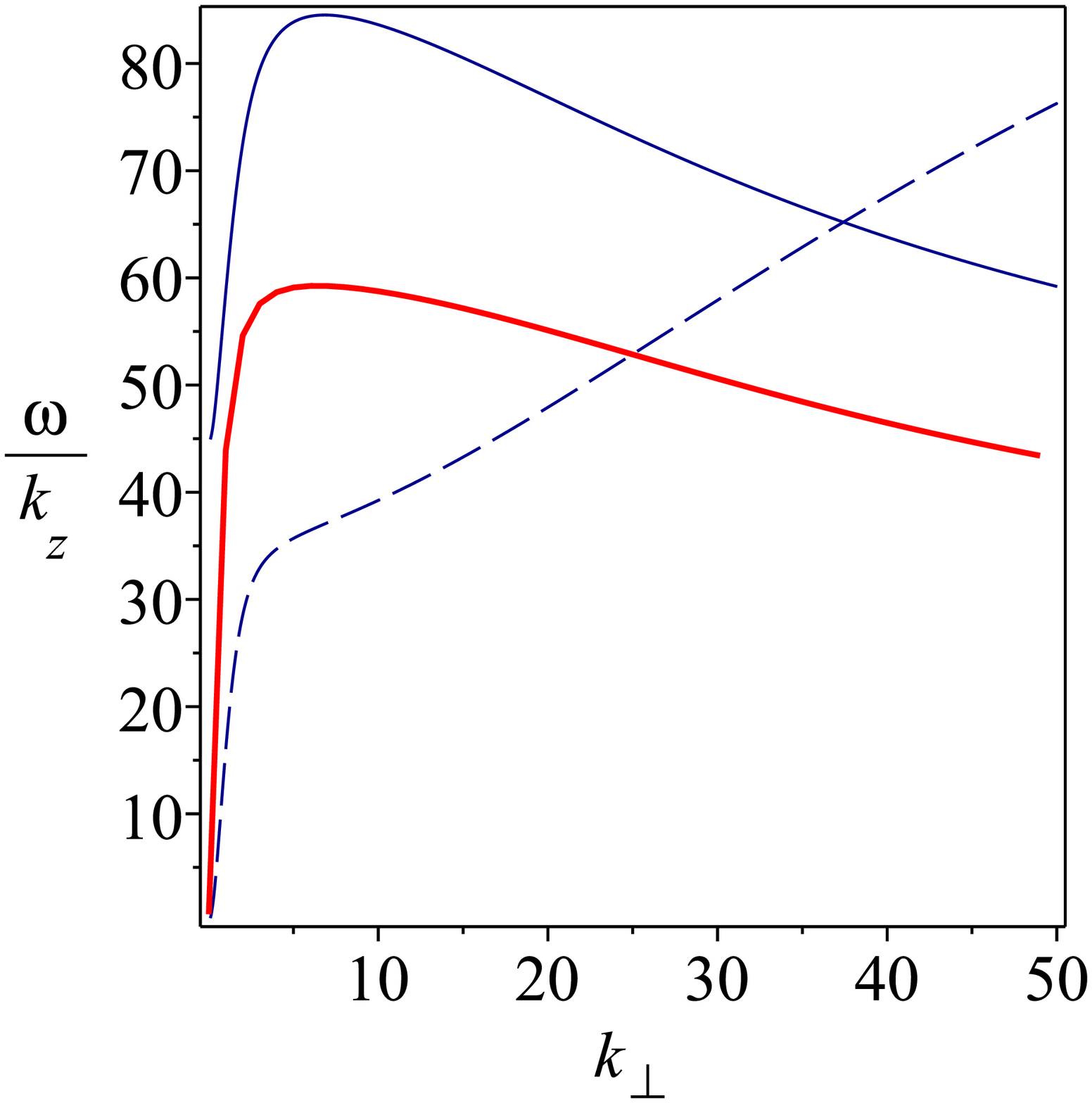}
		\includegraphics[width=0.48\textwidth]{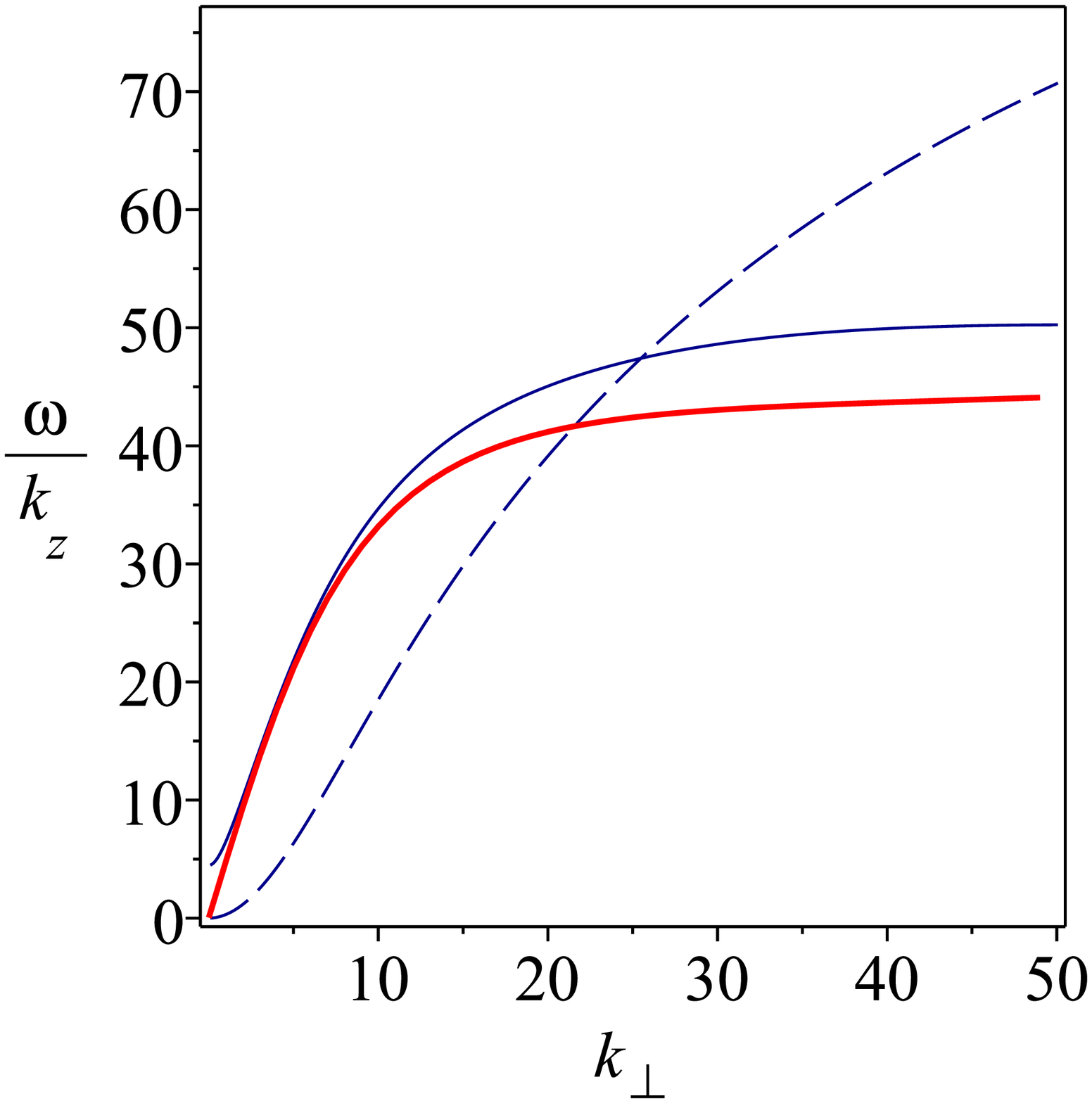}}
	\centerline{
		\includegraphics[width=0.48\textwidth]{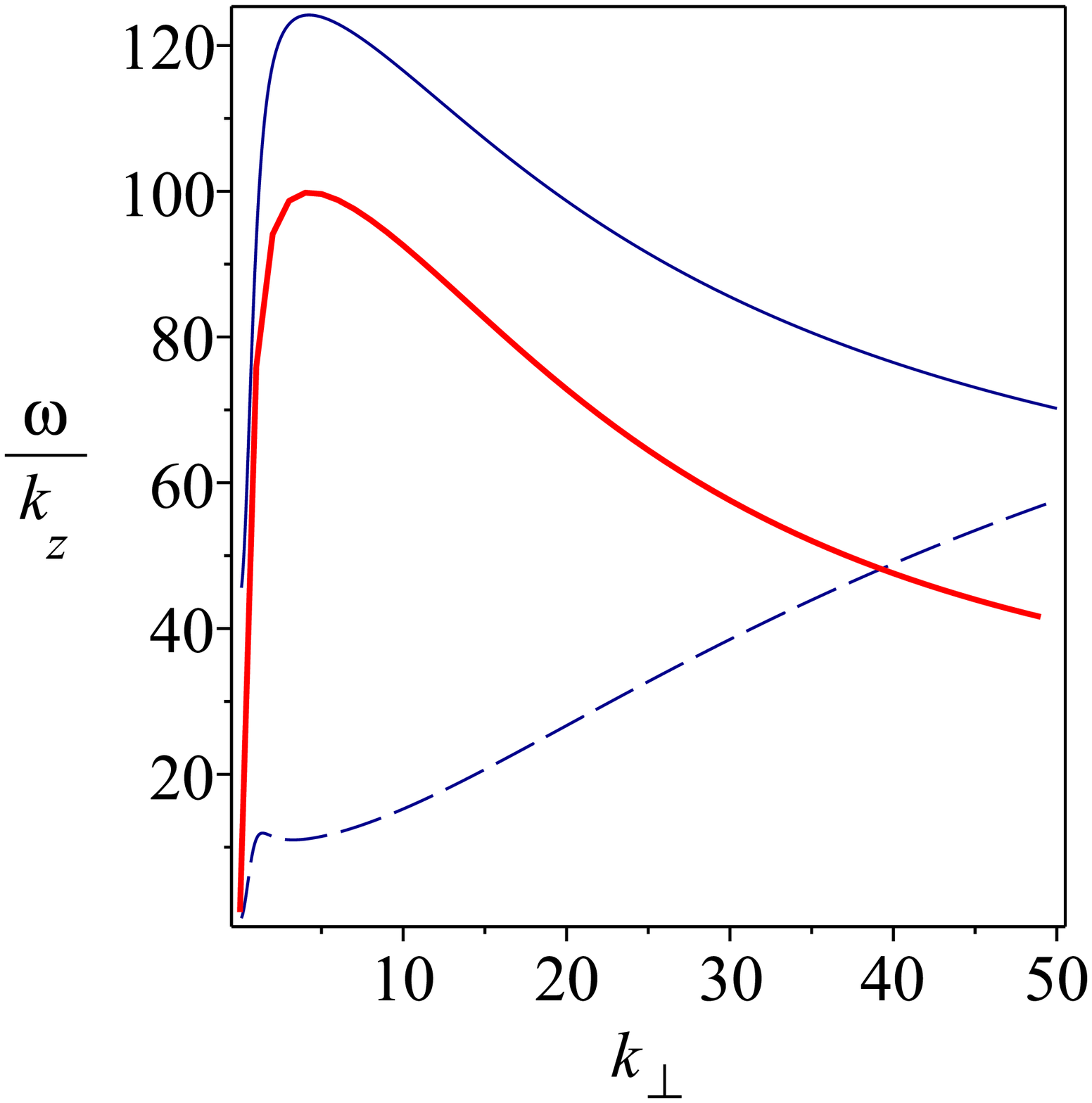}
		\includegraphics[width=0.48\textwidth]{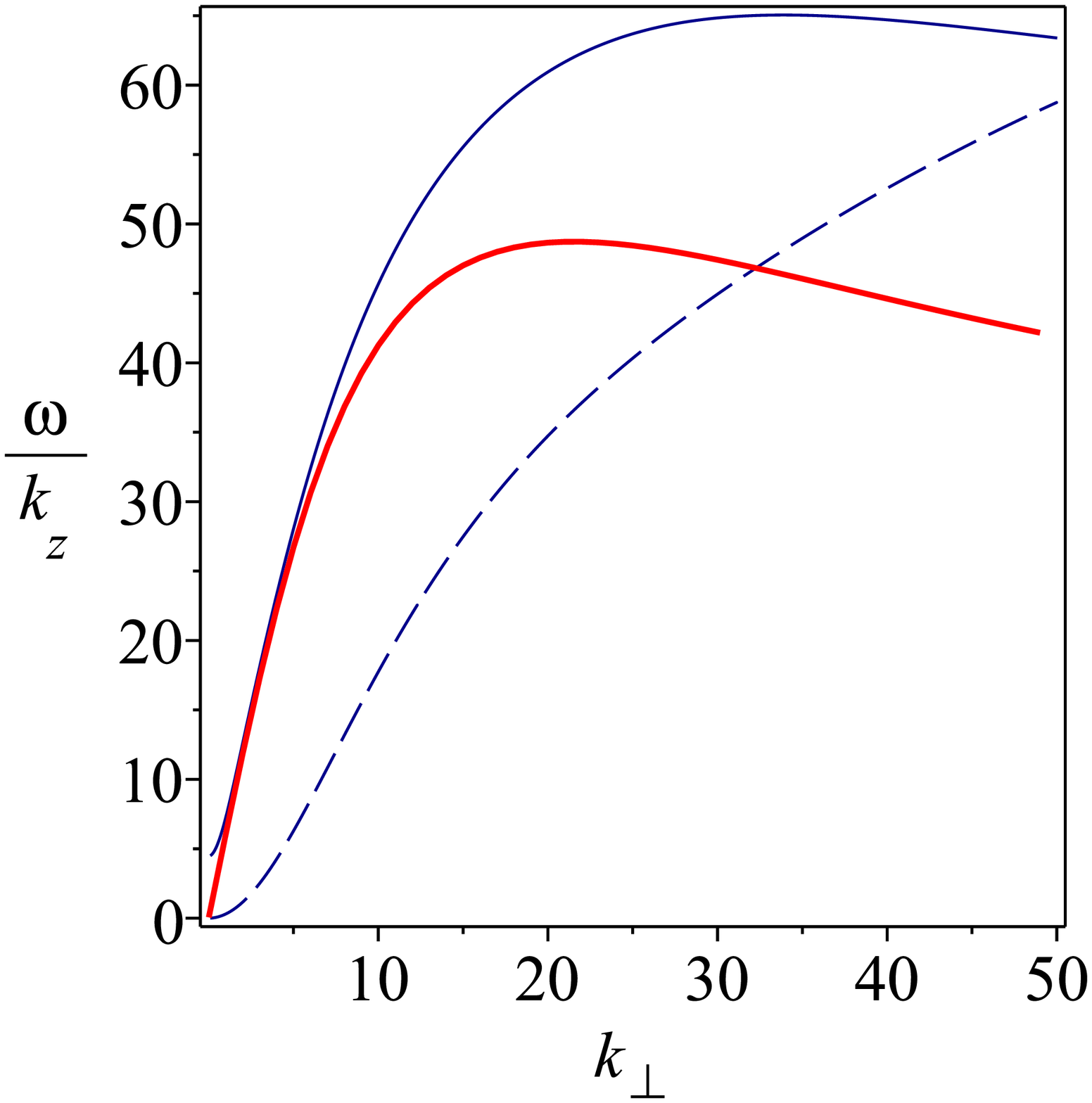}}
	\centerline{
		\includegraphics[width=0.48\textwidth]{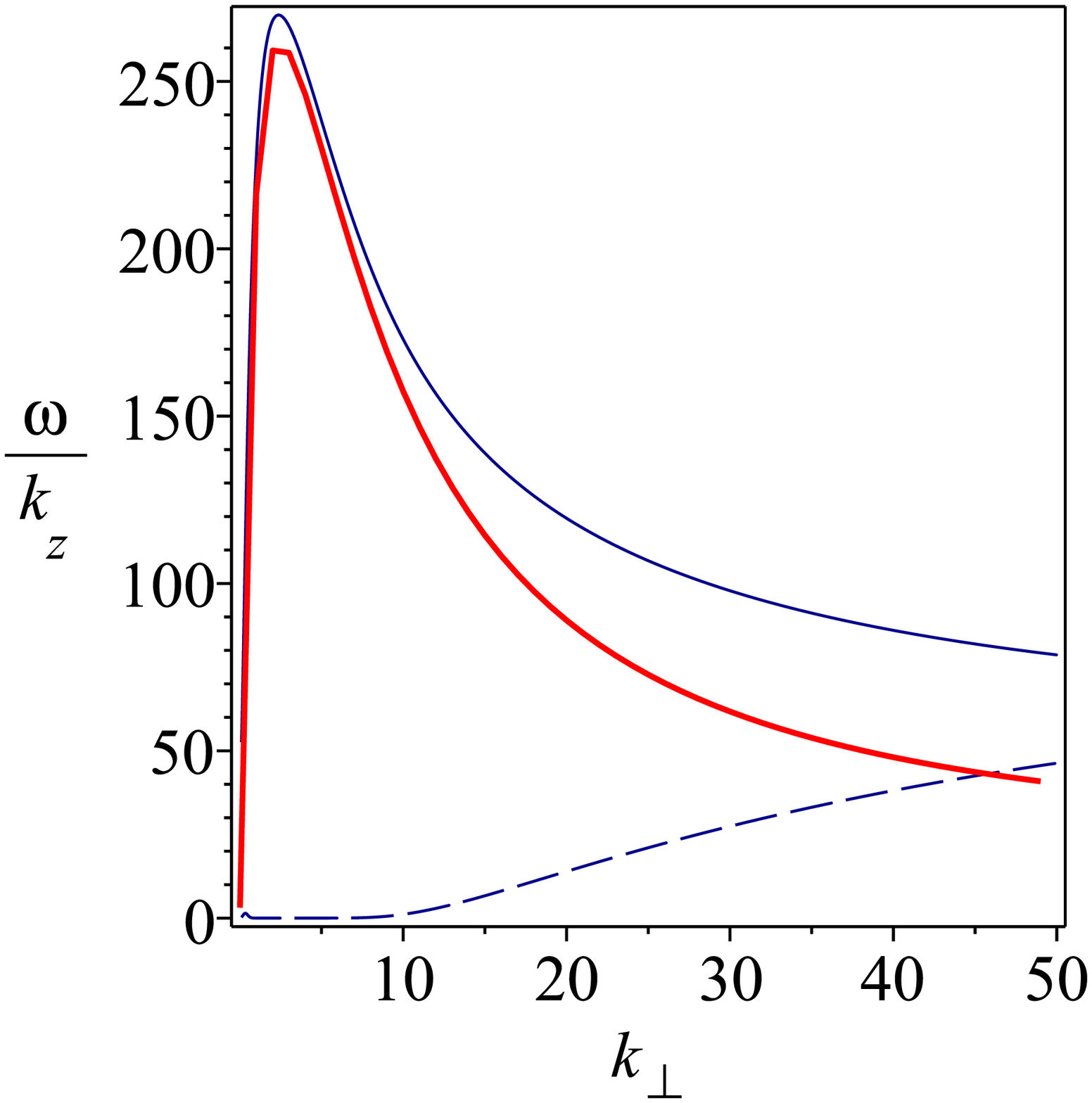}
		\includegraphics[width=0.48\textwidth]{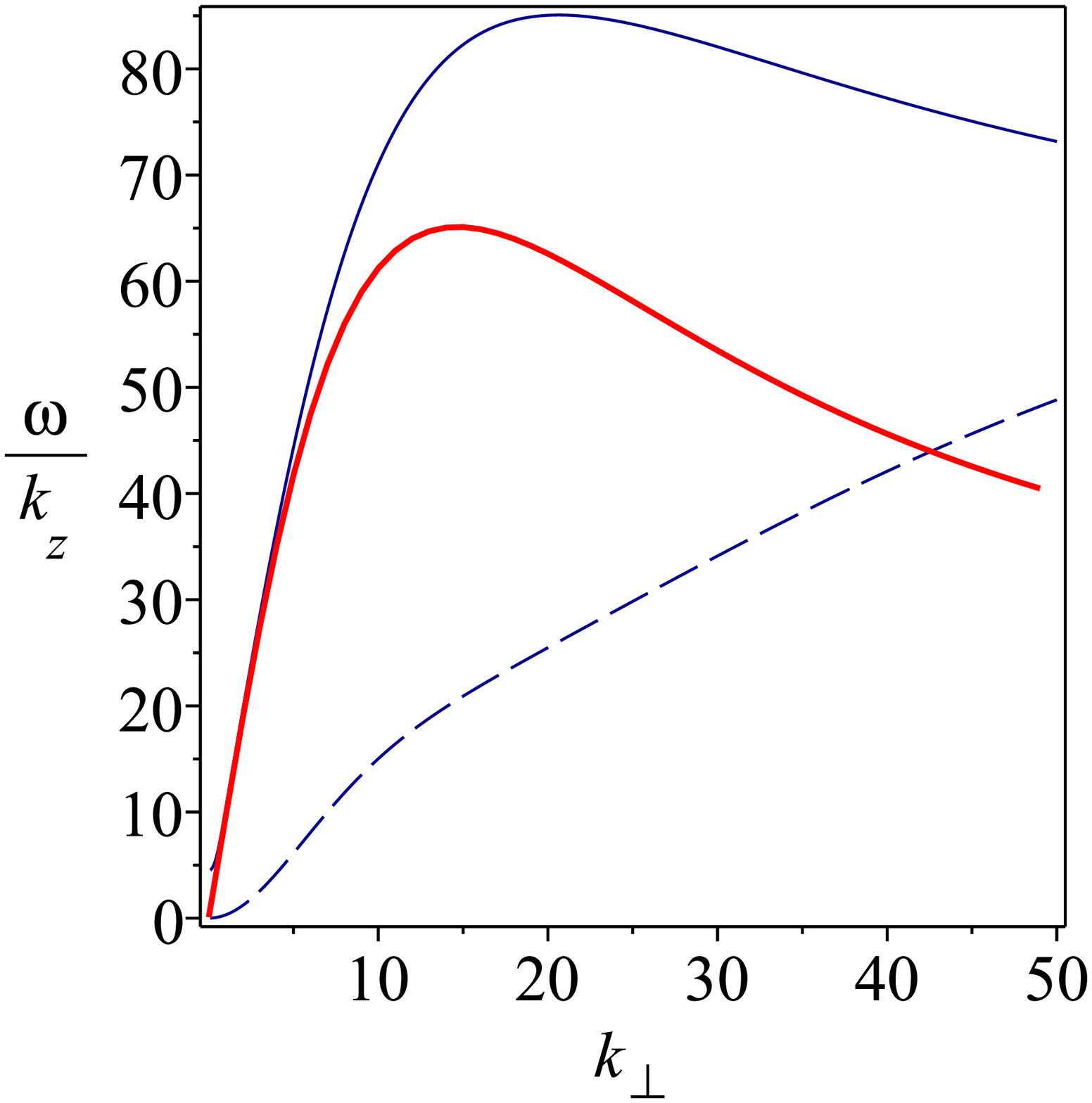}}
	\caption{$\Re(\omega)/k_z$  (solid line) and
		$-\Im(\omega)/k_z$  (dashed line) from kinetic theory (in dark
		blue) and  $\omega/k_z$
		from the two-field model (thick red line) for KAWs at
		$\beta_e=0.001$ (left) for $\tau=1$ (top), $\tau=5$ (middle) and
		$\tau=50$ (bottom), and at $\beta_e=0.1$ (right) for $\tau=0.2$
		(top), $\tau=1$ (middle) and $\tau=5$ (bottom).}
	\label{fig1}
\end{figure}

\subsection{Isothermal regime} 
 Linearizing Eqs. (\ref{eq:A-KAW-iso2})-(\ref{eq:phi-KAW-iso2})
 and (\ref{eq:A-WW-iso2})-(\ref{eq:phi-WW-iso2}), which correspond to the small $\beta_e$ regime, we obtain the  dispersion relation for the KAWs
 \begin{equation}
\frac{\omega}{k_z}=\sqrt{\frac{2}{\beta_e}}\frac{(1+\frac{\delta^2
    k_\perp^2}{2})k_\perp} {\left (1+\frac{2\delta^2k^2_\perp}{\beta_e} +
  \frac{2\delta^4k^4_\perp}{\beta_e} \right )^{1/2}\left
  (\frac{\beta_e}{2}(1+\frac{2\delta^2k^2_\perp}{\beta_e})+\frac{1}{1+\tau}(1-\delta^2k^2_\perp)  \right )^{1/2}},\label{eq:dispersion-KAW}
 \end{equation}
and for the WWs
  \begin{equation}
\frac{\omega}{k_z}=\frac{2}{\beta_e}\frac{(1+\frac{\delta^2
    k_\perp^2}{2})k_\perp} {\left (1+\frac{2\delta^2k^2_\perp}{\beta_e} +
  \frac{2\delta^4k^4_\perp}{\beta_e} \right )^{1/2}\left
  ( 1+\frac{2\delta^2k^2_\perp}{\beta_e} \right )^{1/2}},\label{eq:dispersion-W}
 \end{equation}
respectively, where $k_z$  denotes the 
parallel  wavenumber.
Note that, when $(1+\tau)\beta_e \gg 1$, the dispersion relation of
KAWs and WWs coincide. In that case, when neglecting FLR corrections, the
frequency of the waves saturates at a value
$\displaystyle{\omega_s=\frac{k_z}{k_\perp}\frac{1}{\delta^2}}$,
which, at large enough propagation angle
$\theta_{\boldsymbol{k}\boldsymbol{B}}$, identifies (in dimensional
units) with $\Omega_e \cos \theta_{\boldsymbol{k}\boldsymbol{B}}$. 
On the other hand, when $(1+\tau)\beta_e \ll 1$, KAWs obey, in
the high-frequency domain (still neglecting FLR effects),
$\omega/k_z\simeq (1+\tau)^{1/2}/\delta$.

For both types of waves, the eigenvector is associated with the
relation
\begin{equation}
\widehat{\varphi}^*=\frac{\omega}{k_z}\left (\frac{1+\frac{2\delta^2
		k_\perp^2}{\beta_e}+\frac{2\delta^4
		k_\perp^4}{\beta_e}}{1+\frac{\delta^2}{2}k_\perp^2} \right )\widehat{A_\|},\label{eq:eigenvector}
\end{equation}
where the hat symbol refers to Fourier modes.
From this formula, the magnetic compressibility 
\begin{equation}	
	 C= \frac{{\widehat B}_z^2 }{ |{\widehat {\boldsymbol B}_\perp}|^2} = 
	 \frac{\beta_e^2}{4} \frac{{\widehat \varphi*}^2}
	 {k_\perp^2 {\widehat  A_\|^2}}
\end{equation} 
is easily obtained for each type of wave in the form
\begin{eqnarray}
&& C_{KAW} = \frac{1+ \frac{2\delta^2 k_\perp^2}{\beta_e} + \frac{2\delta^4 k_\perp^4}{\beta_e}}{1 + \frac{2}{(1+\tau)\beta_e} + \frac{\tau}{1+\tau}
\frac{2\delta^2k_\perp^2}{\beta_e}}, \\
&&C_{WW} = \frac{1+ \frac{2\delta^2 k_\perp^2}{\beta_e} + \frac{2\delta^4 k_\perp^4}{\beta_e}}{1 + \frac{2\delta^2k_\perp^2}{\beta_e}}.
\end{eqnarray}
While the magnetic compressibility is of order one for WWs, it displays a sharp increase for the KAWs at scales comparable to $d_e$. Capturing the saturation observed 
in data from the Magnetospheric Multiscale
(MMS) mission, near the electron Larmor radius \citep{Chen-Boldyrev17} nevertheless 
requires a suitable descriptions of the FLR effects at this scale,
which is beyond the scope of the present model.

It is of interest to compare the predictions of the present 
reduced fluid models with
those of the kinetic theory obtained using the WHAMP
software \citep{WHAMP}. The first question concerns the domain of
existence of the two types of wave.
For the KAWs, a propagation angle
$\theta_{\boldsymbol{k}\boldsymbol{B}}$ close to $90^\circ$ is
required in order to prevent the occurrence of cyclotron
resonance within the considered range of wavenumbers.
Note that this condition is relaxed when
$\beta_i=\tau\beta_e$ is large (typically a few units)
\citep{SBG12,PSH12}. For these large values of $\beta_i$, and at
large propagation angles, there is
only one electromagnetic mode at scales smaller than the ion Larmor
radius. It is a continuation of
the shear Alfv\'en branch, and is named Alfv\'en-whistler mode because
its frequency can greatly exceed the ion gyrofrequency \citep{SBG12}. For smaller
$\beta_i$, KAWs and WWs are clearly distinct modes which appear to be the 
continuation of shear Alfv\'en and fast waves respectively. They furthermore
exist in different domains of $\theta_{\boldsymbol{k}\boldsymbol{B}}$ and
$\beta_e$, whistler modes requiring a smaller angle
of propagation (see also \citet{BHXP13}).
If one assumes that the frequency $\omega_{W}$ of WWs obeys  $\omega_{WW} >k_\perp v_{th\, i}> \Omega_i$,
inserting the  approximate (dimensional) formula
$\omega_{WW}\simeq \frac{\Omega_i}{\beta_e}k_z k_\perp \rho_s^2$,
(valid in the range $kd_e<1$), one finds that,
for $k_\perp\rho_s\simeq 1$, one should have $k_z\rho_i\gtrsim \beta_i$, while
for $k_\perp\rho_s=O(1/\mu)=\beta^{1/2}/\delta$, the condition is $k_z/k_\perp\gtrsim
(\tau\beta_e)^{1/2} \delta$.
On the other hand, taking for the KAWs, $\omega_{KAW}\simeq
\Omega_i\beta_e^{-1/2}k_z k_\perp \rho_s^2$, the condition
$\omega_{KAW}<\Omega_i$ taken for $k_\perp
\rho_s\simeq\beta^{1/2}/\delta$ gives the condition $k_z/k_\perp
\lesssim \delta^2\beta_e^{-1/2}$.
In the following, we choose $\theta_{\boldsymbol{k}\boldsymbol{B}} =82^\circ$ for the
WWs and $\theta_{\boldsymbol{k}\boldsymbol{B}} =89.99^\circ$
for the KAWs. 

Figure \ref{fig1} displays the ratios $\Re(\omega)/k_z$ (in dark blue solid
lines) and $-\Im(\omega)/k_z$  (in dark blue dashed lines) obtained from
the linear kinetic theory for KAWs at $\beta_e=0.001$  for $\tau=1$,
$\tau=5$ and $\tau=50$ (left) and at  
$\beta_e=0.1$ for $\tau=0.2$, $\tau=1$ and $\tau=5$
(right). Superimposed in thick red solid lines are the corresponding
ratios $\omega/k_z$ of 
the isothermal dispersion relation given by Eq. (\ref{eq:dispersion-KAW}).
At small $\beta_e$, the agreement between kinetic theory and the
isothermal model is better for large values of $\tau$, in part due to
a smaller damping rate. A much faster decrease of
the ratio $\Re(\omega)/k_z$ is observed for $\tau=50$ and the behavior
is indeed close to that of WWs (see Fig. \ref{fig2},
right), as predicted by inspection of the dispersion relations.
For smaller values of $\tau$ and/or larger values of $\beta_e$, the
damping becomes quite strong when $k_\perp$ reaches a few units.

Turning to the WWs, it is of interest to first briefly
discuss their properties when varying angles and $\beta_e$. As mentioned
previously, for small enough $\beta_i$, WWs can be found as a
continuation to small scales of the fast mode. At a sufficiently
small angle of propagation, e.g. $60^\circ$, they do not encounter
any resonance, even at $\beta_i$ of order unity, but as the angle
and/or $\beta_i$ increases, branches of Bernstein modes cross the
whistler branch, that nevertheless remains continuous throughout the
considered range of wavenumbers, if
$\theta_{\boldsymbol{k}\boldsymbol{B}}$ and $\beta_i$ remain below
certain thresholds. 
This is illustrated in Fig. \ref{fig2} (left), which  displays the
frequencies (solid) and damping rates (dashed) at 
  $\beta_e=0.01$, $\tau=8$ and $\theta_{\boldsymbol{k}\boldsymbol{B}} =82^\circ$ for
  an Alfv\'en wave (red), a fast wave ending in the first Bernstein
  mode (blue) and in the second one (brown) and for the whistler
  wave (green) continuing to small scale (as a red dotted
  line in the right panel). 
Choosing $\theta_{\boldsymbol{k}\boldsymbol{B}} =82^\circ$ and
$\beta_e=0.01$, we display in the right panel of Fig. \ref{fig2},
$\Re(\omega)/k_z$ (solid lines) and $-\Im(\omega)/k_z$ (dashed lines) for
WWs at $\tau=0.2$ (dark blue), $\tau=1$
(brown), $\tau=5$ (green) and $\tau=8$ (red dots), superimposed with
the prediction of 
linear theory (Eq. (\ref{eq:dispersion-W})) (thick black line). As
$\tau$ increases, the kinetic results  
converge to the same curve which is very close to the
prediction of the model, as long as dissipation remains small
(i.e. for $k_\perp \le 10$). A very similar behavior is observed for a
$75^\circ$ propagation angle.

Left panel of Fig. \ref{fig3} displays the electron density fluctuation
of the eigenmode, obtained from the kinetic theory for KAWs (red
line) at $\theta_{\boldsymbol{k}\boldsymbol{B}} =89.99^\circ$ and
WWs (dark blue line) at 
$\theta_{\boldsymbol{k}\boldsymbol{B}} =82^\circ$ for  $\beta=0.01$ and
$\tau=5$. This result, which  remains true at larger values
of $\beta_e$, confirms the assumption that was used to
distinguish the two waves, namely that the WWs are associated
with almost incompressible motions.

\begin{figure}
  \centerline{
\includegraphics[width=0.48\textwidth]{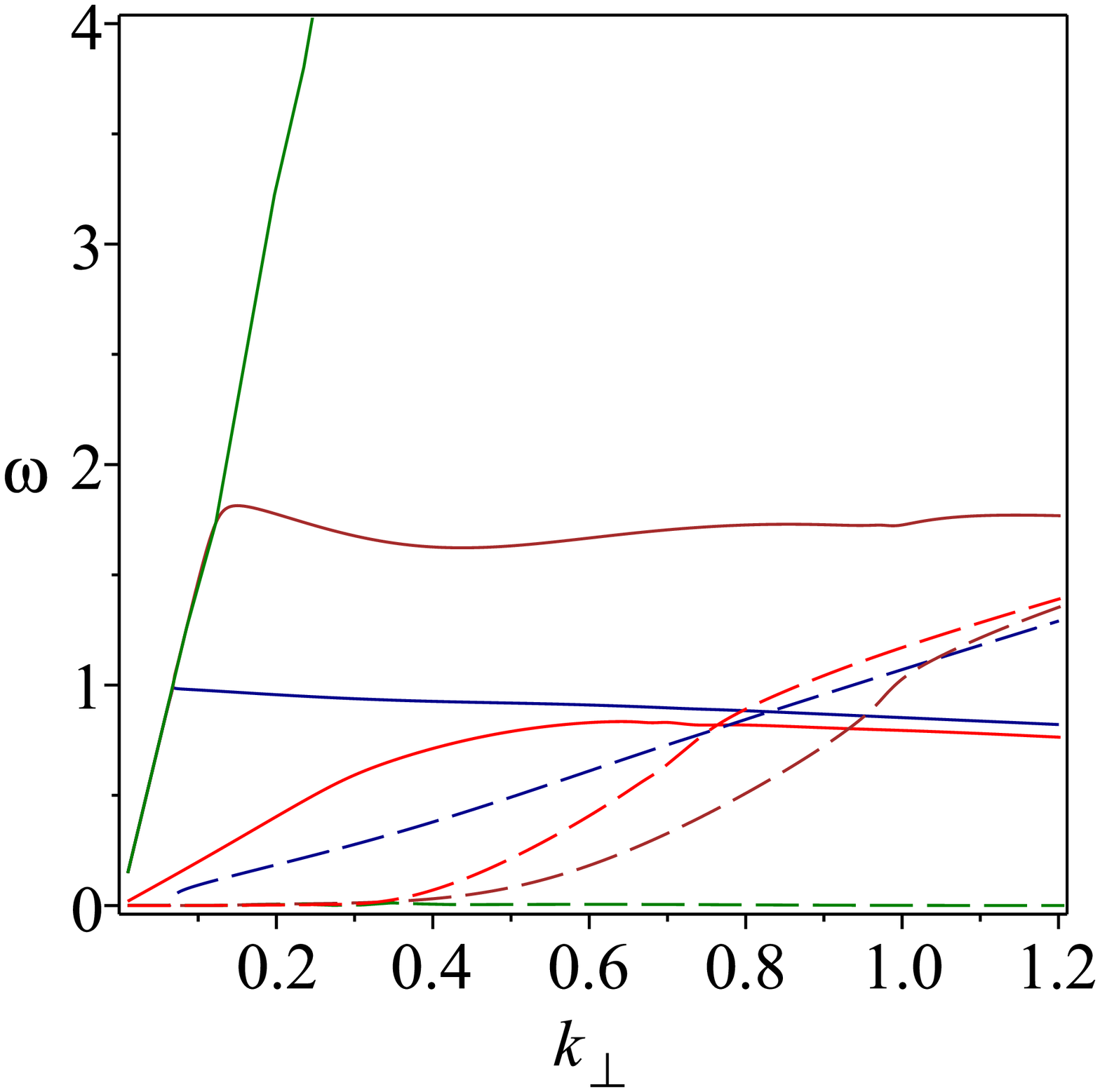}
\includegraphics[width=0.48\textwidth]{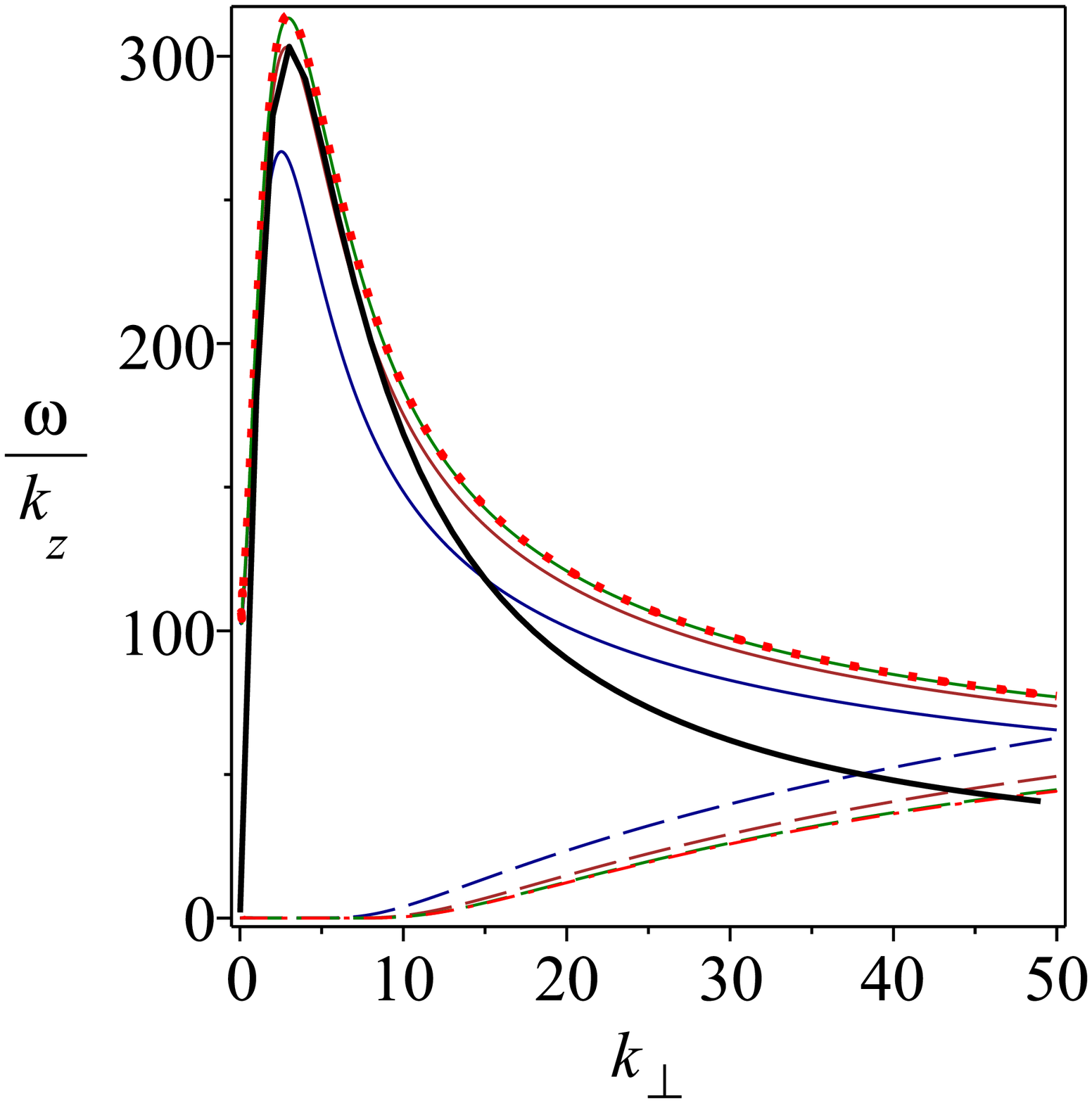}}
\caption{Left:  frequencies (solid) and damping rates (dashed) at
  $\beta_e=0.01$, $\tau=8$ and $\theta_{\boldsymbol{k}\boldsymbol{B}} =82^\circ$ for
  an Alfv\'en wave (red), a fast wave ending in the first Bernstein
  mode (blue) and in the second one (brown) and for the whistler
  wave (green) continuing to small scale (see right in red dotted
  line).
  Right:  $\Re(\omega)/k_z$
  (solid lines) and $-\Im(\omega)/k_z$
  (dashed lines) from kinetic theory 
  for whistlers at $\beta_e=0.01$ for $\tau=0.2$ (dark blue),
  $\tau=1$ (brown), $\tau=5$ (green) and $\tau=8$ (red dotted)  and
  ratio $\omega/k_z$  from the two-field model (thick black line).}
\label{fig2}
\end{figure}

\begin{figure}
\centerline{
\includegraphics[width=0.48\textwidth]{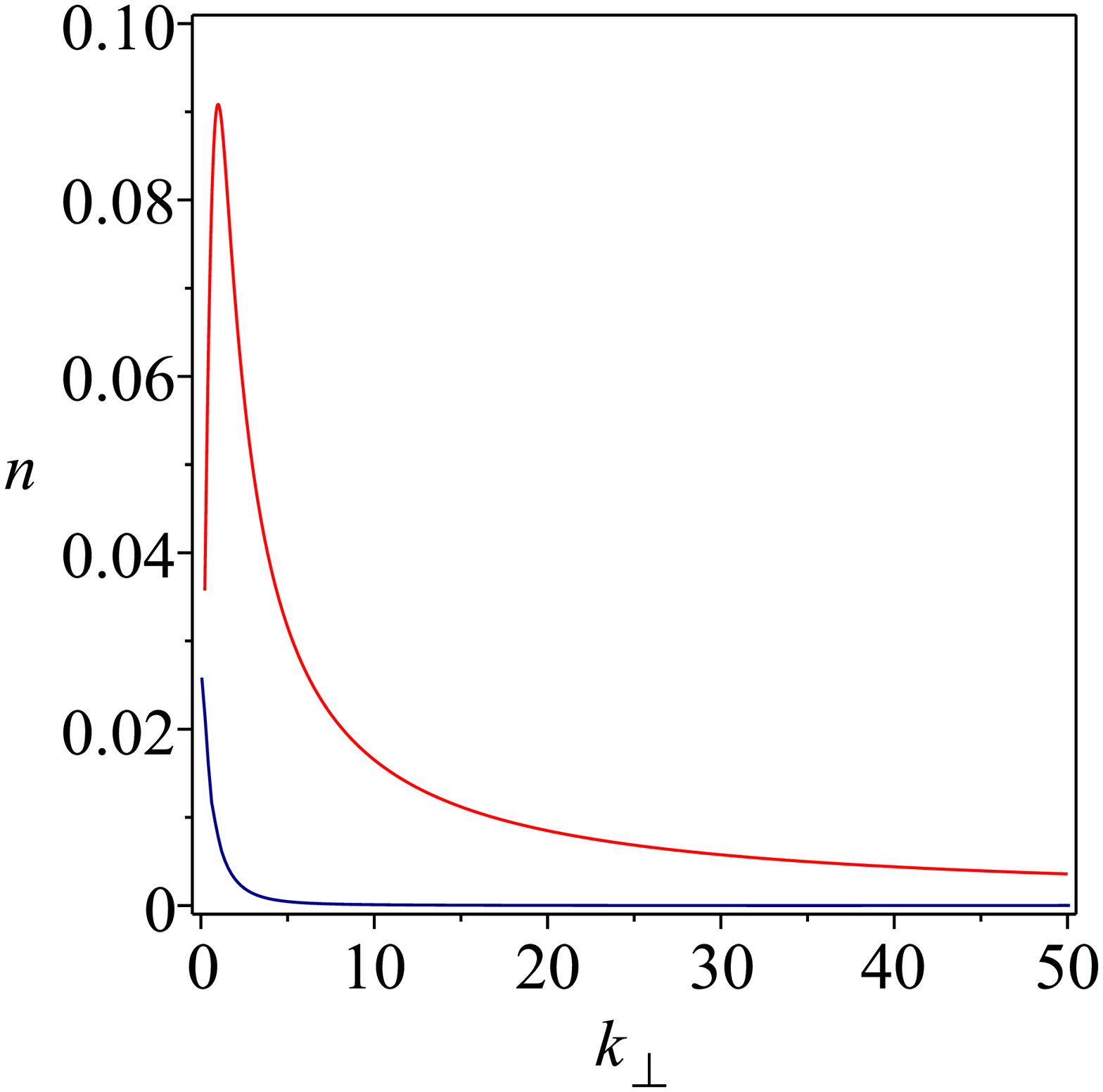}
\includegraphics[width=0.48\textwidth]{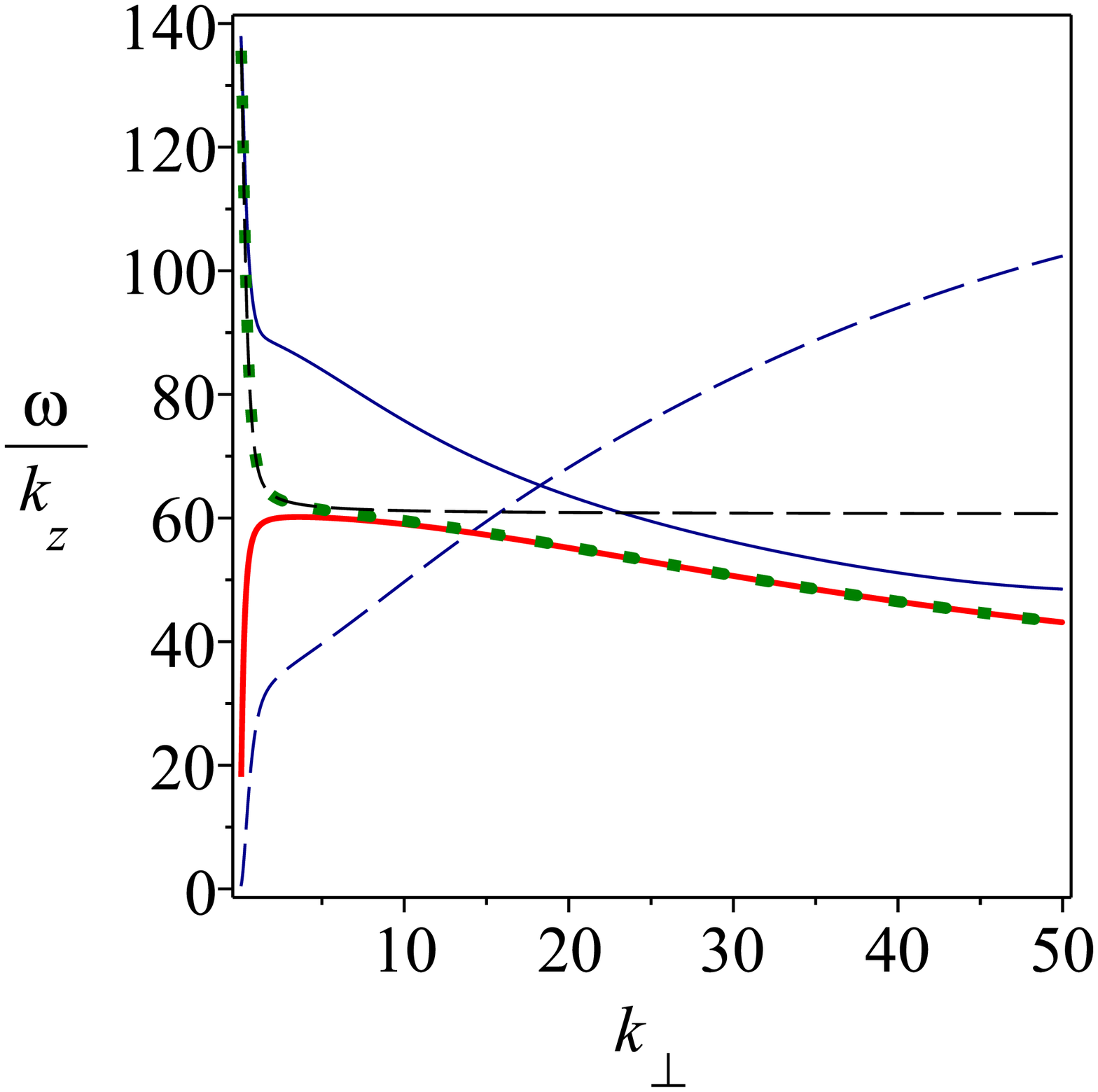}}
\caption{Left: electron density fluctuations for whistlers at
  $\theta_{\boldsymbol{k}\boldsymbol{B}} =82^\circ$ (dark blue) and 
  KAWs  at   $\theta_{\boldsymbol{k}\boldsymbol{B}} =89.99^\circ$
  (red) for $\beta_e=0.01$ and $\tau=5$.  Right: $\Re(\omega)/k_z$ for a KAW
  at $\beta_e=0.0001$ and $\tau=1$, from
  kinetic theory (in dark blue) and frequency 
  from various two-field models: from Eq. (\ref{eq:dispersion-KAW})
  (thick red line), from the 
model of \citet{ZS11} taken in the isothermal limit (dashed black
line), and from an 
extension of the current model using the improved relation between
$n$ and $\varphi$ given by Eq. (\ref{eq:gamma0}) (green dotted line). The dark blue
dashed line displays $-\Im(\omega)/k_z$.}
\label{fig3}
\end{figure}

Finally, the right panel of  Fig. \ref{fig3} displays the case of
a smaller $\beta_e$  for KAWs, a situation where the behavior at small 
$k_\perp$ is now quite sensitive to the ion response. We display
$\Re(\omega)/k_z$ for  $\beta_e=10^{-4}$ and $\tau=1$ for the kinetic
theory (dark blue solid line) and different models. The prediction
of Eq. (\ref{eq:dispersion-KAW}), which is displayed with a thick red solid
line, shows a strong disagreement for $k_\perp <5$ . The agreement with kinetic theory is however much better at these large scales when using the dispersion relation 
of the model of \citet{ZS11} taken in the isothermal limit (black dashed line). Note that 
the different curvature of the dispersion relation, observed 
at small $k_\perp$ when $\beta_e$ crosses $m_e/m_i$, is associated with the well-known transition from KAWs to inertial Alfv\'en waves. 
Interestingly, when modified  by using the gyrokinetic Poisson 
equation  (\ref{eq:gamma0}) (green dots), our model reproduces for
$\omega/k_z$ both the qualitative behavior at  large scales and the
decrease at small scales, in agreement with kinetic theory, in spite of a different absolute level in a spectral range where the Landau damping is nevertheless so strong that the corresponding waves are rapidly dissipated.

\subsection{Reduced Landau fluid model}

\begin{figure}
	\centerline{
		\includegraphics[width=0.48\textwidth]{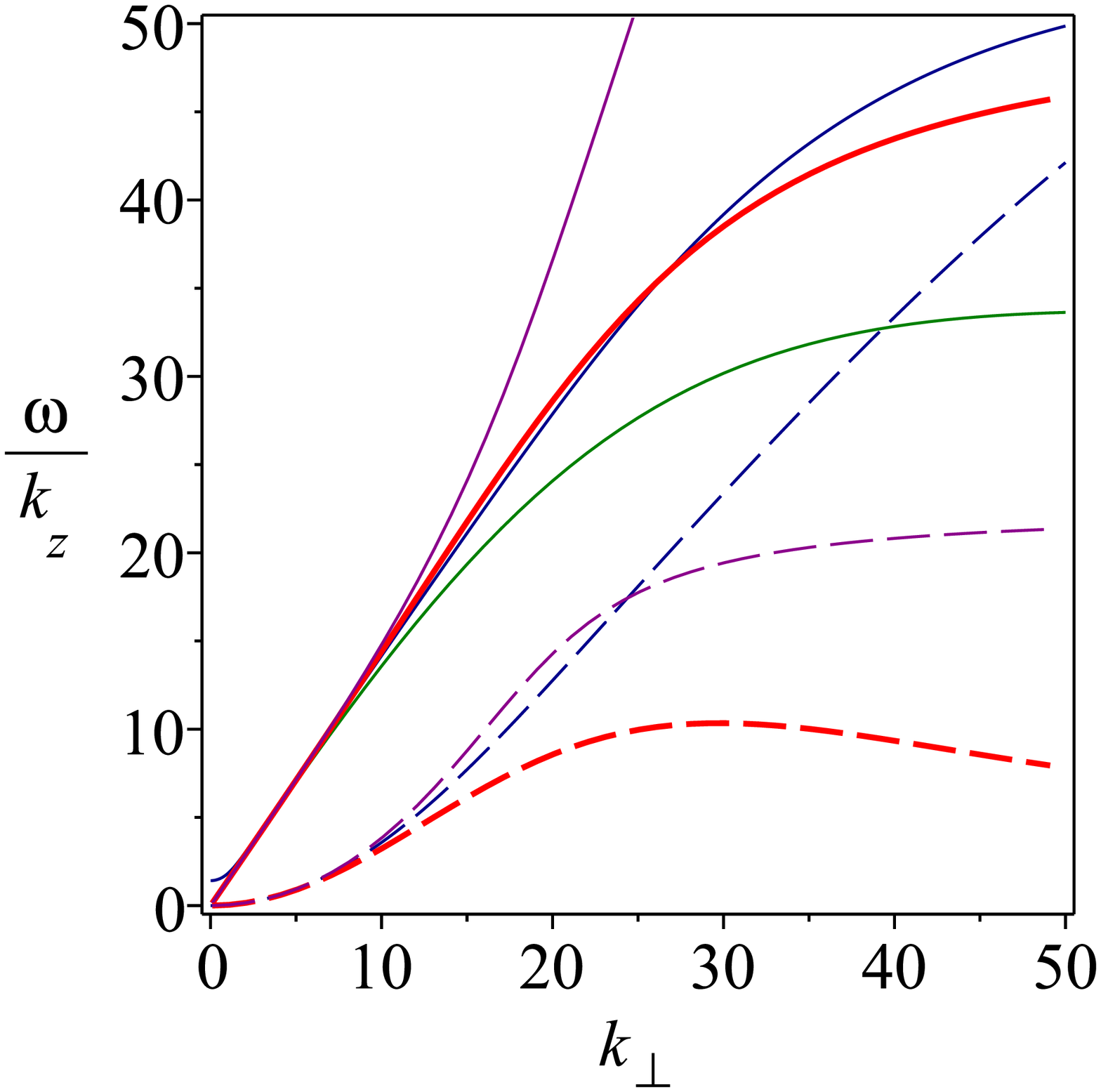}
		\includegraphics[width=0.48\textwidth]{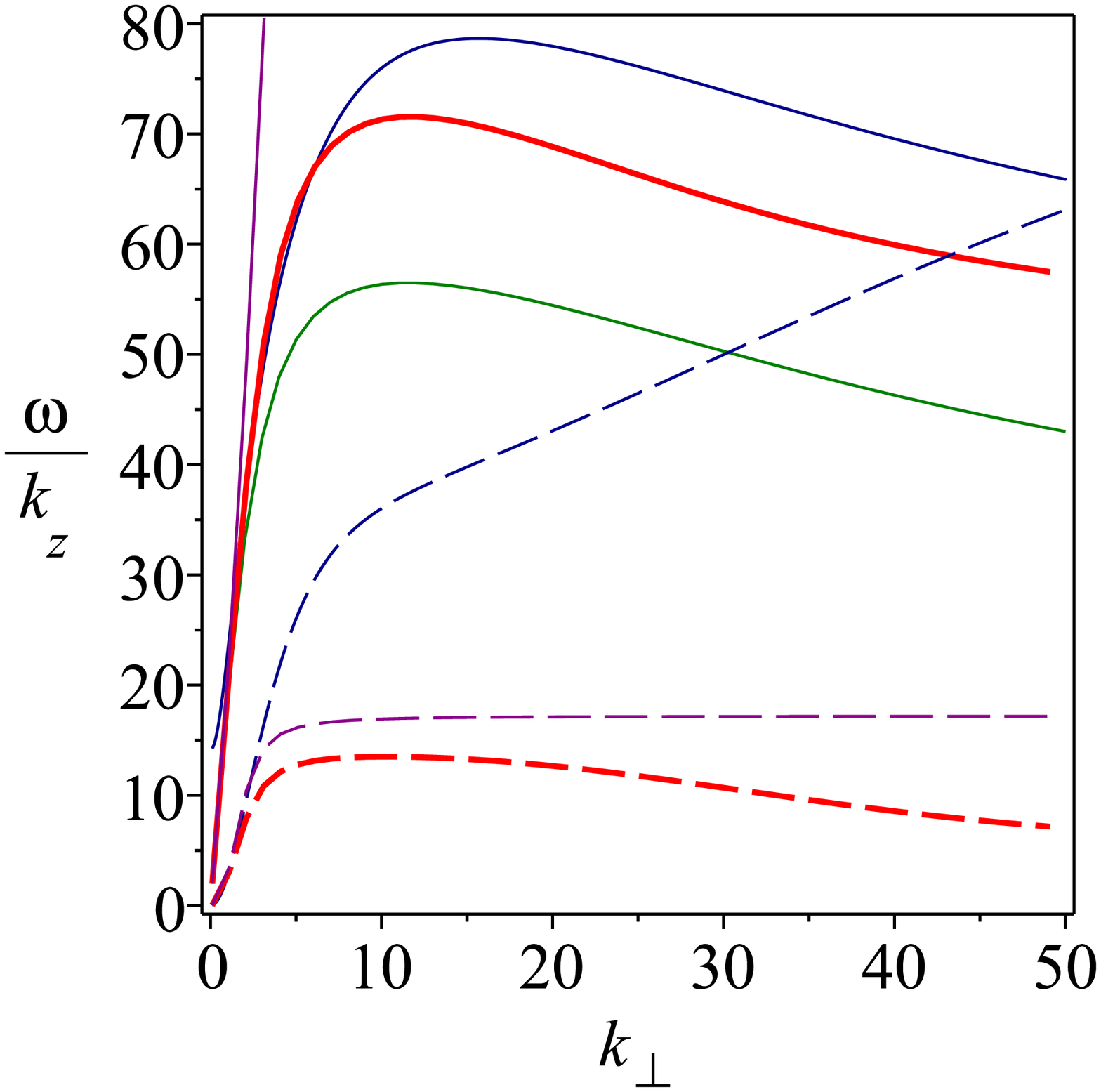}
	}
	\caption{$\Re(\omega)/k_z$  (solid lines) and $-\Im(\omega)/k_z$ (dashed lines) 
		for KAWs with propagation angle $\theta_{{\boldsymbol k}
		{\boldsymbol B}	}= 89.99^\circ$, at $\tau=1$ and $\beta_e=1$ (left) or $\beta_e=0.01$ (right) 
		from kinetic theory  (dark blue),
		the RLF model (thick red lines), the isothermal
		two-field model (green) and the RLF model with $\delta=0$ (violet).}
	\label{fig4}
\end{figure}

We shall here restrict the discussion to the impact of Landau
damping on the dispersion relation for KAWs.
Equations ({\ref{eq:A-unif})-(\ref{eq:Tperp2}) supplemented by Eqs. ({\ref{eq:qparal})-(\ref{eq:rperp}),
are linearized,  and the dispersion
relation computed using the MAPLE software. Figure \ref{fig4}  displays
$\Re(\omega)/k_z$  (solid lines) and $-\Im(\omega)/k_z$
  (dashed lines) for KAWs at $\tau=1$, in both cases  $\beta_e=1$
(left) and $\beta_e=0.01$ (right), for different reduced models, with
the kinetic theory prediction superimposed in dark blue. The green
and thick red lines correspond to the isothermal and RLF models
respectively. A clear extension of the spectral validity range is
obtained when Landau damping is retained. The
damping rate is however globally weaker than predicted by the
kinetic theory, except at the largest scales, 
possibly pointing out the limitation of the closure assumption at
small scales. It turns out that this discrepancy is not significantly reduced by
 retaining dynamical equations for the gyrotropic fluxes and closing the fluid 
hierarchy at the next order. The agreement is also not improved when instead of a
 closure assumption we use the low-frequency kinetic formulas given in Appendix B of
\citet{PS07} and take for the plasma response functions Pad\'e approximants 
of various orders. This observation raises the question whether Landau damping is 
the only dissipation mechanism acting at these scales.

 Noticeably for
$\beta_e=1$, the comparison with kinetic theory for the real part is
satisfactory up to $k=25$ (very close to the inverse electron Larmor
radius). In the case where electron inertia and FLRs are
not included (violet line), a regime where no saturation is
expected, the disagreement starts around $k=10$. This latter model, which
can be viewed as an extension of the model of \cite{BHXP13}
including Landau damping, 
was originally derived in \citet{TSP16} (where the heat fluxes were
estimated directly from the linear kinetic theory, instead of using the
above approximate closure). 
For $\beta_e=0.01$,  globally similar graphs are obtained except
that, saturation occurring at smaller wavenumbers, the domain where
the kinetic theory is accurately reproduced is more limited.
  
\section{Turbulent regimes} \label{sec:turb}

In this section, we concentrate on  the isothermal models, 
considered in the spectral ranges ${\mathcal D}_e^< = \{1/\rho_s \ll k_\perp \ll 1/d_e\}$
and ${\mathcal D}_e^> = \{1/d_e\ll k_\perp\ll 1/\rho_e \}$ separately.  While in the former domain,
where $\delta^2k_\perp^2/\beta_e \ll 1$, electron inertia is negligible, this effect is dominant 
in the latter, where $\delta^2k_\perp^2/\beta_e \gg 1$.
In both ranges, FLR effects are subdominant corrections that we will thus neglect.
The energy becomes
\begin{equation}
  {\mathcal E}_{KAW} \propto \frac{1}{2}\int\left (\frac{1}{1+\tau}\varphi^{*2}
  + \frac{2}{\beta_e}|\bnabla_\perp A_\||^2 +
  \frac{4\delta^2}{\beta_e^2}(\Delta_\perp A_\|)^2\right ) d{\boldsymbol x},
\end{equation}
for the KAWs, and 
\begin{equation}
  {\mathcal E}_{WW}  \propto  \frac{1}{2}\int  \left (\varphi^{*2} + \frac{2\delta^2}{\beta_e} |\bnabla_\perp \varphi^*|^2+
  \frac{4}{\beta_e^2} |\bnabla_\perp A_\||^2 +
  \frac{8\delta^2}{\beta_e^3} (\Delta_\perp A_\|)^2\right ) d{\boldsymbol x},
\end{equation}
for the WWs. In more physically explicit terms, we can write
\begin{equation}
  {\mathcal E}_{KAW} \propto \frac{1}{2} \int \left (  \frac{2}{\beta_e} {|\boldsymbol B}_\perp|^2
   +  (\tau +1 ) n^2 +\delta^2 u_{\|e}^2   \right ) d{\boldsymbol x}
\end{equation}
and
\begin{equation}
  {\mathcal E}_{WW}  \propto  \frac{1}{2} \int \left (\frac{2}{\beta_e}  |{\boldsymbol B}|^2 +\delta^2  |{\boldsymbol u}_e|^2  \right)   d{\boldsymbol x}.
\end{equation}
Note that these energies are dominated by the transverse magnetic energy at the largest scales and by the
parallel kinetic energy at the smallest ones. In a turbulent regime, the energy is expected to cascade 
to small transverse scales, resulting in an inertial range
characterized by a scale-independent energy flux $\epsilon$. The aim of this section is to phenomenologically evaluate
the scaling properties  of the transverse magnetic energy spectrum $E_{B_\perp}(k_\perp)$ 
in such a cascade, in both regimes of strong and weak turbulence, assumed to exist and to
result from three-wave interactions. 

When assuming transverse isotropy, the energy  flux $\epsilon$ can
 be  estimated as $k_\perp$ times  the one-dimensional spectral
  density, divided by the characteristic transfer time $\tau_{tr}$ at
   this  wavenumber. The regimes of
strong and weak turbulence differ in  the estimate of this time that,
 in both cases,  can nevertheless be written 
 $\tau_{tr} = \tau_{NL}^2/ \tau_{L}$, where $\tau_{NL}$ refers to the characteristic time
of the nonlinear interactions at the corresponding scale, and $\tau_L\sim 1/\omega$ 
to the inverse frequency of the considered wave at this scale. 

In the strong turbulence regime,
where a so-called critical balance holds, the rate of the nonlinear interaction at a given scale is comparable
to the frequency of the linear wave at this scale ($\tau_{NL} \sim \tau_L$), leading us to identify  $\tau_{tr}$ with $\tau_{NL}$. 
In this approach, $\tau_{NL}$ is defined as the shortest of the characteristic times 
associated with the various nonlinear couplings, which turns out to be be given by
$\tau_{NL}^{-1} \sim [\widehat\varphi^*, \cdot]$ associated with the transverse strain. Note that the same estimate of the transfer time holds in the absence of waves.
We are thus led to write
$1/\tau_{NL} \sim k_\perp^2 {\widehat \varphi^*}$. An additional element used 
to relate the magnitude of the various fields is provided by the condition that the considered solution is an eigenmode of the
linear problem, thus obeying equation (\ref{eq:eigenvector}). 
In both spectral ranges defined above, one can write the phase velocity in the asymptotic form  $\omega/k_\| \sim k_\perp^\alpha$.
In ${\mathcal D}_e^<$, one thus has ${\widehat \varphi^*} \sim (\omega/k_\|) {\widehat A_\|}$, 
leading to $\tau_{NL}^{-1} \sim k_\perp^{1+\alpha} {\widehat B}_\perp$, while  in ${\mathcal D}_e^>$,
${\widehat \varphi^*} \sim (\omega/k_\|) k_\perp^2 {\widehat A_\|}$, 
leading to $\tau_{NL}^{-1} \sim k_\perp^{3+\alpha} {\widehat B}_\perp$.

In the strong turbulent regime, it follows that in
${\mathcal D}_e^<$, where  $\epsilon \sim  k_\perp^{1+\alpha} {\widehat B_\perp}^3$, one gets
$E_{B_\perp}(k)\sim \epsilon^{2/3} k_\perp^{-(5+2\alpha)/3}$, while in ${\mathcal D}_e^>$, where
$\epsilon \sim  k_\perp^{5+\alpha} {\widehat B_\perp}^3$, one has $E_{B_\perp}(k)\sim \epsilon^{2/3} k_\perp^{-(13+2\alpha)/3}$.

In the weak turbulent regime, one easily obtains that in ${\mathcal D}_e^<$, $\tau_{tr}^{-1} \sim k_\perp^{(2+\alpha)}B_\perp^2$,
while in ${\mathcal D}_e^>$, $\tau_{tr}^{-1} \sim k_\perp^{(6+\alpha)}B_\perp^2$. It follows that in ${\mathcal D}_e^<$,
$E_{B_\perp}(k)\sim \epsilon^{1/2} k_\perp^{-(2+\alpha/2)}$, while in ${\mathcal D}_e^>$,
$E_{B_\perp}(k)\sim \epsilon^{1/2} k_\perp^{-(5+\alpha/2)}$.
Let us now consider more specifically the cases of KAWs and WWs.

In ${\mathcal D}_e^<$, the dispersion relation is the same for KAWs and WWs, 
with $\alpha = 1$. As a consequence, for both types of waves, $E_{B_\perp}(k_\perp)\sim \epsilon^{2/3} k_\perp^{-7/3}$ in the
strong turbulent regime and $E_{B_\perp}(k_\perp)\sim \epsilon^{1/2} k_\perp^{-5/2}$ in the weak turbulent regime,
a result obtained as the finite flux solution of the weak turbulence spectral equations in
\citet{Galtier03}. 

In  ${\mathcal D}_e^>$,
various cases are to be distinguished. For the KAWs, when the ion-to-electron temperature ratio $\tau$ is moderate,
$\alpha =0$.  One thus obtains  $E_{B_\perp}(k_\perp)\sim \epsilon^{2/3} k_\perp^{-13/3}$ in the strong turbulent regime
and  $E_{B_\perp}(k_\perp)\sim \epsilon^{1/2} k_\perp^{-5}$ in the
weak turbulent regime. Note that a $-13/3$ exponent  for the
magnetic field 
spectra has also  been reported in the different context of a 
generalized helicity cascade in incompressible so-called extended
MHD \citep{Krishan04,Abdelhamid16}. In contrast, for the WWs, but also for the
KAWs when the ion-to-electron temperature ratio is large ($\tau\gg 1$), $\alpha =-1$. It follows that in this case 
$E_{B_\perp}(k_\perp)\sim \epsilon^{2/3} k_\perp^{-11/3}$ in the
strong turbulence regime  and $E_{B_\perp}(k_\perp)\sim
\epsilon^{1/2}k_\perp^{-9/2}$ in the weak turbulence regime.
A similar $k_\perp^{-11/3}$ spectrum for KAWs 
at large $\tau$ is also obtained by \citet{Chen-Boldyrev17}.

Furthermore, in the case of strong turbulence, the spectral anisotropy can be estimated from the above relations. While at large scale, the balance between the 
linear term $\partial_z \varphi^*$ in Eq. (\ref{eq:A-KAW-iso2}) 
with the bracket $[\varphi^*, A_\|]$ leads to 
$k_\| \sim k_\perp^2{\widehat A_\|}$ and thus to the usual scaling
$k_\|\sim k_\perp^{1/3}$, at small scales the relevant nonlinear 
term is $[\varphi^*, (2\delta^2/\beta_e)\Delta_\perp A_\|]$. In this regime, $k_\| \sim k_\perp^4 {\widehat A_\|}$ where 
${\widehat A_\|} \sim k_\perp^{-8/3-\alpha/3}$. For KAWs with 
$\tau \sim 1$, $\alpha=0$, this implies  $k_\| \sim k_\perp^{4/3}$, 
while for WWs or KAWs at $\tau \gg 1$, one has $k_\| \sim k_\perp^{5/3}$, consistent with the finding of \citet{Chen-Boldyrev17}. In both cases, 
the longitudinal transfer becomes increasingly important at smaller scales.	 

It is also of interest to consider the effect of a finite value of $\beta_e$ on the energy
spectrum. Considering the case of KAWs, Fig.  \ref{fig3} (right) shows that 
taking $\beta_e = 10^{-4}$ causes
$\omega/k_\|$  to decrease with $k_\perp$ (red solid and green dotted lines),
while it tends to a constant when $\beta_e = 0$ (black dashed line). As the spectrum can be written in the form
$k_\perp^{-13/3} (\omega/k_\|)^{-2/3}$, this suggests that a finite $\beta_e$ will tend to make the spectrum shallower.

At the numerical level, simulations of whistler turbulence were performed using 
incompressible bi-fluid equations in 2.5D (i.e. three-dimensional velocities and magnetic fields
but two-dimensional space coordinates), showing 
a transition from a $-7/3$ spectrum at the ion scales to a steeper slope (close
to $-11/3$) at scales smaller than $d_e$ \citep{Andres14}.
Three-dimensional particle-in-cell simulations of WW turbulence also  clearly
indicate the development of two sub-ion spectral ranges, but the exponents
are steeper than those predicted by the above phenomenology, and
 non-universal \citep{Chang11,Gary12}.
Kinetic effects, such as electron Landau damping could be at the origin of this discrepancy.
This issue could be addressed with the RLF model (discussed in Section \ref{sec:landau}),
both numerically and using a phenomenological approach as in \citet{PS15}. 
Concerning the kinetic Alfv\'en wave cascade, 
the phenomenology discussed above assumes a transformation of Alfv\'en 
waves to KAWs near the ion gyroscale. Recent numerical simulations 
\citep{CerriNJP17} however indicate that fast magnetic reconnection
processes can inject energy on a time scale shorter than the nonlinear cascade time  at the reconnection scales, leading to the formation
of $d_i$-scale structures and to a nonlinear cascade that fills the spectrum at small scales. The effect of such phenomena on the sub-ion turbulence remains to be analyzed quantitatively, possibly using reduced fluid models.

\section{Conclusion} \label{sec:conclu}
Reduced fluid models have been derived for the sub-ion-scale
dynamics of collisionless plasmas, retaining electron inertia and leading-order electron FLR
corrections. Neglecting the ion dynamics, we are led to discriminate
between two limiting cases: a low-frequency regime
involving perpendicular pressure balance and a high-frequency one,
where the electron fluid is essentially incompressible. 
The two resulting models capture KAWs and WWs respectively, in the
context of isothermal or Landau fluid closures.
They extend the validity range of previously existing models. At small $\beta_e$ and 
scales large compared to $d_e$, they both reduce to ERMHD
\citep{Schekochihin09,BHXP13} where the same equations govern
different fields depending on the kind of waves.
Furthermore, at scales comparable to $d_e$, in 2.5D dimensions
 and for  $\beta_e$ small enough  for
electron FLR corrections to be negligible (i.e. $\beta_e=O(m_e/m_i)$), 
the system for WWs reproduces 
EMHD (see e.g. Eqs. (10)-(11) of \citet{Biskamp99}). Differently,
in the same $\beta_e$-range, the equations for KAWs with the Landau fluid
closure appear as a fluid reduction of 
those of \citet{ZS11}  (see \cite{Loureiro13} for their numerical
simulations in the case of fast collisionless reconnection). While the latter
model is based on a drift kinetic description of the electrons, the
present RLF model includes first-order FLR corrections. Such terms,
together with the contribution of parallel magnetic fluctuations,
which are both $O(\beta_e)$, induce a sensitivity of the system to
pressure anisotropy, an effect potentially important in a reconnection context
\citep{Le16}.

Computation of the electron gyroviscous force
$\bnabla \bcdot  \boldsymbol{\Pi}_e$ is
performed within an asymptotic expansion based on temporal and
spatial scale separation between the electron gyroradius and the
considered quasi-transverse scales, a procedure which, at scales comparable to
$d_e$, results in an expansion in terms of  $\beta_e$.
The computation involves a recursive process involving both  the
non-gyrotropic pressure and heat flux tensors, two iterations being needed in order
to obtain the leading order in $\beta_e$. Because of the algebraic
complexity in the general framework (see e.g. \citet{Ramos05a}), we
resorted in the present  paper to assume a gyrotropic heat flux
in the expression of the gyroviscous force.
Relations with previous FLR estimates is discussed in Appendix \ref{append:classical}. Note that
including the gyroviscous force in a generalized Ohm's law may in particular be
useful for enriching hybrid models in the context of collisionless reconnection.

Both isothermal and Landau fluid closures are considered for 
KAWs and WWs. The linear regime is examined in comparison with
the fully kinetic theory to evaluate the validity of these
closures. In the isothermal case, a  qualitative agreement is
found, which turns out to be more accurate when Landau damping is weak.
For the propagation angles exceeding $80^\circ$ that we have considered, WWs
are more weakly damped than KAWs, at least at scales larger than
$\rho_e$. As expected, for the values of $\beta_e \lesssim 0.1$ 
and moderate values of $\tau$, the range of angles where KAWs can propagate without
resonance turns out to be restricted to quasi-perpendicular angles.
The Landau fluid closure improves the model accuracy. In
particular, when $\beta_e$ is pushed to values of order unity, the
range of wavenumbers where the dispersion relation is accurately described is significantly enlarged.

Observations in the solar wind and the magnetosheath 
provide evidence of power-law magnetic energy spectra at scales
smaller than $d_e$, possibly associated with turbulent
cascades. Assuming that KAWs and WWs are still present at such
scales, we explored the 
spectrum of the transverse magnetic fluctuations  
in both  strong and weak turbulence energy cascades.
As a first step, a phenomenological approach where FLR
corrections are neglected is presented.  In addition to the
well-known  WWs magnetic spectra both above and below $d_e^{-1}$, and to
the $-7/3$  sub-ion KAWs spectrum, a new regime of KAW
turbulence is obtained at scales smaller than  $d_e$,  characterized by a
$-13/3$ exponent for strong turbulence and by a $-5$ exponent in the weak
regime. These exponents are to be compared with satellite
observational data that display slopes steeper than the WWs $-11/3$
spectrum. Note that other physical effects such as Landau damping
\citep{PS15,Sulem2016}) 
and intermittency corrections associated with coherent structures, such as 
current sheets \citep{BHXP13},
can also lead to steeper spectra. It should be stressed that both
phenomenological arguments and numerical simulations predict a
transition at $d_e$, while observational spectra
in the terrestrial magnetosheath only display a transition at $\rho_e$ \citep{Huang14},
associated with electron demagnetization. In fact, a transition at $d_e$ is
more clearly observed at the level of the magnetic compressibility 
\citep{Chen-Boldyrev17}. This does exclude that  additional effects 
such as  other kind of waves or  structures could play a role.

As shown in Section \ref{sec:closures}, our four-field model, as well as its two-field isothermal reductions, 
conserve energy. 
To the best of our knowledge, this is the first example of reduced fluid model for
inertial reconnection conserving energy and accounting for electron FLR effects.
A question we intend to address in the future concerns the Hamiltonian structure of this model.
Previous results concerning the inclusion of (ion) FLR effects in Hamiltonian reduced fluid models
were presented in \cite{Mor84c,Haz87,Dag05,Mor14b,Iza11}. It could in particular be of interest to investigate whether
the idea of the gyromap transformation, adopted in \cite{Mor84c,Haz87,Iza11,Mor14b}, together with  the Hamiltonian
structure of the model in the absence of FLR contributions, can help in building a Hamiltonian model including  electron FLR effects.
Further developments should also include numerical
simulations to test the predictions for the turbulent energy spectra,
and in particular to evaluate the role of coherent
structures and  analyze the difference between two- and
three-dimensional geometries. Existence and stability of Alfv\'en or 
electron vortices \citep{Mikhal87,Schep94} are also open questions.
Other open issues
concern the relative importance of KAWs and WWs
in space plasmas. Such questions cannot be addressed using  gyrokinetic
simulations or gyrofluid models which, involving a  perpendicular pressure balance,
concentrate on low-frequency waves. Furthermore, the models for the WWs and the KAWs
considered in this paper differentiate at the level of the determination of the
magnitude of the density fluctuations. When the latter are neglected, the
system can be viewed as an extension of EMHD for WWs to smaller scales, while when
pressure balance is prescribed, it describes KAWs dynamics. It would
be of great interest to derive a reduced fluid model able to  simultaneously capture
the two types of wave. Noticeably , recent hybrid kinetic simulations indicate that the relative importance of KAWs and WWs is sensitive to the plasma $\beta$, KAWs being dominant at $\beta \gtrsim 1 $ and WWs at lower $\beta$ \citep{CerriApJ16}.
	
In addition to turbulent cascades discussed in
this paper, an important issue that the present models can address
concerns collisionless magnetic reconnection, in particular in three
dimensions where fully kinetic simulations  require huge
computational resources. In the turbulent regime, RLF models could in
particular provide an efficient tool to study the relative contributions  of 
coherent structure disruptions  and incoherent fluctuation cascades  in the processes
of dissipation and plasma heating \citep{Parashar15}, an important issue aimed at 
being addressed by the  THOR satellite mission \citep{THOR16}.

\appendix

\section{Nongyrotropic electron pressure tensor} \label{append:FLR-pressure}
As derived in \cite{Scheko10}, an exact equation for the electron
pressure tensor $\mathsfbi{P}_{e}$ reads, when neglecting collisions,
\begin{eqnarray}
&&\mathsfi{P}_{e,ij}=p_{\perp e}{\delta}_{ij}+(p_{\perp
    e}-p_{\| e})
  \hatb_i\hatb_j +\delta^2\frac{\mathsfi{M}_{ijkl}}{4B} \Big [
  \frac{D^{(e)}}{Dt}\mathsfi{P}_{e,kl} +\partial_m
  \mathsfi{Q}_{e,mkl} \nonumber\\
  &&+ \left ({\delta}_{mn}\mathsfi{P}_{e,kl}  +
  {\delta}_{kn}\mathsfi{P}_{e,ml}+{\delta}_{ln}\mathsfi{P}_{e,mk}
       \right)\partial_m { u}_{e,n} \Big ].\label{eq:press-tensor}
\end{eqnarray}
In this equation, $\mathsfbi{Q}_{e}$ denotes the heat flux tensor
and
\begin{equation}
  \mathsfi{M}_{ijkl}=\left (
  {\delta}_{ik}+3\hatb_i\hatb_k \right )
  {\epsilon}_{jln}\hatb_n +
   {\epsilon}_{iln}\hatb_n \left ( {\delta}_{jk}+3\hatb_j\hatb_k
  \right ).
\end{equation}
Equation (\ref{eq:press-tensor}) can be solved recursively, using
as small parameters $\delta^2 \omega$ and $\delta^2 k^2$,
corresponding in the dimensional variables to the conditions that 
the considered scales $k^{-1}$ be large and the frequencies $\omega$
small compared to the electron Larmor radius and cyclotron frequency
respectively. For $\beta_e$ of order unity, this also prescribes
that scales must be large compared to $d_e$. In this regime, the
appropriate scaling is that given by Eqs. (3.13) of
\citet{TSP16}. Differently, for small values of
$\beta_e$, scales comparable to $d_e$ can be considered, the
appropriate scaling being that given in Section \ref{sect:Faraday}.
Keeping the leading-order corrections in $\beta_e$ nevertheless requires in this
case to also retain the second order relatively to the scale
expansion, as shown below. In the following, we shall concentrate on
the small $\beta_e$ regime, but the resulting gyroviscous force
remains valid at $\beta_e=O(1)$ (recalling that, in the present units,
$k$ must remain small compared to
$\displaystyle{(\frac{m_i}{2m_e})^{1/2}\approx 30}$),
although it then includes subdominant contributions.

The iteration mentioned above involves in fact the coupling with another equation for
the heat flux tensor (not written here). Solving the coupled system
proves to be quite involved and falls outside 
the scope of the present paper. A linear version can be found in
\cite{GPS05}.
Here, for the sake of simplicity, 
we resorted to only retain the contribution of the gyrotropic part
of $\mathsfbi{Q}_{e}$ in the evaluation of the gyroviscous stress.

\subsection{First-order contributions}
At zeroth order, the pressure tensor is simply given by its
gyrotropic expression $\mathsfbi{P}^G_{e}=p_{\perp
  e}\mathsfbi{I}+(p_{\perp e}-p_{\| e})\boldsymbol{\tau}$.
At first order, replacing $\mathsfbi{P}_{e}$  by $\mathsfbi{P}^G_{e}$
in the rhs of (\ref{eq:press-tensor}), we
easily get for the gyroviscous tensor
\begin{eqnarray}
&&\boldsymbol{\Pi}^{(1)}_e= -\frac{\delta^2}{4B} \Big [ {\widehat
    {\boldsymbol b}} {\boldsymbol \times}  {\mathsfbi W}   
  \bcdot \left ({\mathsfbi I} + 3 \, \boldsymbol{\tau}\right ) 
-\left ( {\mathsfbi I} + 3 \, \boldsymbol{\tau}\right )\bcdot  {\mathsfbi W} 
{\boldsymbol \times} {\widehat{\boldsymbol b}} \Big ] \nonumber  -\frac{\delta^2}{B}\Big
            [ {\widehat{\boldsymbol b}}{\boldsymbol\otimes} 
({\boldsymbol w }{\boldsymbol \times} {\widehat{\boldsymbol b}}) + 
({\boldsymbol w }{\boldsymbol \times} {\widehat{\boldsymbol
      b}}){\boldsymbol\otimes} {\widehat{\boldsymbol b}}\Big ].\nonumber\\ \label{Pi-nl}
\end{eqnarray}
Here,
\begin{equation}
 {\mathsfbi W}  =\left [ p_{\perp e} \bnabla {\boldsymbol u}_e +
   \bnabla (q_{\perp e} \, \bhatb)\right]^S 
\end{equation}
and
\begin{eqnarray}
&&{\boldsymbol w} = (p_{\perp e} -p_{\| e})\Big (\frac{d{\widehat{\boldsymbol b}}}{dt}  + 
\nabla_\|{\boldsymbol u}_e \Big) + (3 q_{\perp e} - q_{\| e}) \nabla_\|{\widehat{\boldsymbol b}},
\end{eqnarray}
where, for a given tensor $\mathsfbi{T}$, the notation
$\mathsfbi{T}^S$ denotes the sum of the tensor with the ones
obtained by circular permutation of the indices.
Assuming an equilibrium state with
isotropic temperatures, the term $w$ is of order $O(\varepsilon^3)$ in the present
asymptotics and can thus be neglected.
We thus write
$\displaystyle{\boldsymbol{\Pi}^{(1)}_e=\boldsymbol{\Pi}^{(1,u)}_e+\boldsymbol{\Pi}^{(1,q)}_e}$ where
\begin{equation}
 \boldsymbol{\Pi}^{(1,u)}_e = -\frac{\delta^2}{4B} \Big [ {\widehat{\boldsymbol b}}
   {\boldsymbol \times} \left [ p_{\perp e} \bnabla {\boldsymbol u}_e\right]^S  \bcdot
   \left ({\mathsfbi I} + 3 \, {\boldsymbol \tau}\right ) -\left ( {\mathsfbi I}
   + 3 \,{\boldsymbol \tau}\right )\bcdot 
\left [ p_{\perp e} \bnabla {\boldsymbol u}_e\right]^S    {\boldsymbol \times}
      {\widehat{\boldsymbol b}} \Big ], \label{eq:PIe}\\
\end{equation}
and where $\displaystyle{\boldsymbol{\Pi}^{(1,q)}_e}$ and all the
quantities derived from it can obtained from the formula involving
$\displaystyle{\boldsymbol{\Pi}^{(1,u)}_e}$ after the replacements
$p_{\perp e}$ by  $1$ and $\boldsymbol{u}_e$ by $q_{\perp e} \bhatb$.
Equation (\ref{eq:PIe}) 
identifies with Eq. (C1) of \citet{Ramos05b}, up 
to the minus sign due to the electron charge.

\subsubsection{Velocity contributions}
Carrying out
calculations similar to those performed in \citet{Ramos05b}, we obtain,
without neglecting any possibly subdominant contribution,
\begin{eqnarray}
 \bnabla\bcdot {\boldsymbol \Pi}^{(1,u)}_e=&&\overbrace{-\delta^2 \left [ \bnabla
   \btimes \left ( \frac{p_{\perp e}} {B}\bhatb \right )
   \bcdot\bnabla \right ] {\boldsymbol u}_e}^{\circone[1]}
 +\overbrace{\frac{\delta^2}{2} \bnabla
 \left (\frac{p_{\perp e} \omega_{\|e}}{B}
 \right )}^{\circone[2]} \nonumber\\
 && \overbrace{+\delta^2 \bnabla \btimes\left (\frac{p_{\perp e}} {B} \nabla_\|
 {\boldsymbol u}_e \right )}^{\circone[3]}
 \overbrace{+\frac{\delta^2}{2} \bnabla \btimes \left ( 
 (\bnabla \bcdot {\boldsymbol u}_e) \bhatb \right )}^{\circone[4]}
 \overbrace{- \frac{3\delta^2}{2}\bnabla \btimes \left (  \bhatb\bcdot \left
 ( \nabla_\|  {\boldsymbol u}_e \right ) \bhatb   \right )}^{\circone[5]}
 \nonumber\\
 && \overbrace{-3\delta^2 B\nabla_\| \left (
 \frac{p_{\perp e}} {B^2}   \bhatb\btimes
 \nabla_\| {\boldsymbol u}_e \right )}^{\circone[6]}
 \overbrace{-\frac{3\delta^2 }{2}B \nabla_\| \left (
 \frac{p_{\perp e}} {B^2}   \omega_{\|e}\bhatb \right )}^{\circone[7]}
 \overbrace{+\delta^2 B \nabla_\| \left (
 \frac{p_{\perp e}} {B^2}  {\boldsymbol \omega}_{e}\right
 )}^{\circone[8]},\nonumber\\
 \label{eq:divPI}
\end{eqnarray}
where $\boldsymbol{\omega}_e=\bnabla\btimes \boldsymbol{u}_e$.
 
Let us first calculate $\bhatb\bcdot \bnabla\bcdot {\mathsfbi \Pi}^{(1,u)}_e$,
which enters Eq. (\ref{Aeq}) where all  the terms scale
as  $\delta\varepsilon^2$ for the KAWs or $\beta_e^{1/2} \delta \varepsilon^2$ for the WWs
(when the scaling parameter $\mu$ is replaced by $\delta/\beta_e^{1/2}$).
Thus, in Eq. (\ref{eq:divPI}), only contributions of these orders,
possibly up to factors of order $O(\beta_e)$
(which is assumed to be only ``moderately'' small) are to be kept.
It follows that the contributions of the terms $\mbox{\circone[4]}$ - $\mbox{\circone[6]}$,
which scale like $\varepsilon^3$ (since in particular
$\bnabla \bcdot {\boldsymbol u}_e$ is of second order in $\varepsilon$
for the KAWs and even smaller for the WWs), are  negligible. The relevant contributions are:
from term $\mbox{\circone[1]}$, $\delta^2 [p_{\perp e}-B_z, u_{\|e}]$;
from term $\mbox{\circone[2]}$, $(\delta^2/2)\nabla_\| \omega_{ze}$;
from term $\mbox{\circone[3]}$,  $\delta^2\bhatz\bcdot\bnabla
  {\boldsymbol\times}  \nabla_\| \boldsymbol{u}_{\perp e}$;
from  term $\mbox{\circone[7]}$,  $-(3\delta^2/2)\nabla_\|\omega_{ze}$;
from  term $\mbox{\circone[8]}$, $\delta^2\nabla_\|\omega_{ze}$.
We thus obtain
\begin{equation}
\bhatb\bcdot \bnabla\bcdot {\boldsymbol \Pi}^{(1,u)}_e=\delta^2\Big \{
             [p_{\perp e}-B_z, u_{\|  e}] +\bhatz\bcdot\bnabla
  {\boldsymbol\times}  \nabla_\| \boldsymbol{u}_{\perp
    e} \Big \}.\label{eq:divPI_par}
\end{equation}
Note that, in terms of $\varphi^*$,  we
can write
\begin{equation}
  \bhatz\bcdot\bnabla
  {\boldsymbol\times}  \nabla_\| \boldsymbol{u}_{\perp
    e}=\nabla_\|\Delta_\perp \varphi^*+\sum_{i=x,y}
  [\partial_i\varphi^*,\partial_i A_\|].\label{eq:phis-expression}
\end{equation}
Furthermore, the term $ \delta^2[p_{\perp e}-B_z, u_{\|  e}]$, which participates
to the so-called gyroviscous cancellation is $O(\delta
\varepsilon^2)$ for the KAWs, thus larger than all
the other contributions, which are $O(\beta_e \delta
\varepsilon^2)$. Differently, for the WWs, 
all the terms are $O(\beta^{3/2} \delta \varepsilon^2)$.

It is also easy to obtain the transverse component of
$\bnabla\bcdot {\boldsymbol \Pi}^{(1,u)}_e$, which reads
\begin{eqnarray}
  ( \bnabla\bcdot {\boldsymbol \Pi}^{(1,u)}_e)_\perp=&&\delta^2\Big \{ [p_{\perp
        e}-B_z,\boldsymbol{u}_{\perp e}]+\frac{1}{2}\bnabla_\perp \left (
  \frac{p_{\perp e}}{B}\omega_{\|e}\right ) \nonumber \\
 && +\frac{1}{2}\bnabla_\perp
   {\boldsymbol\times}\left [ \left ( \bnabla\bcdot
    \boldsymbol{u}_{e}-\nabla_\| u_{\|e} \right ) \bhatz \right
  ] +\nabla_\|\bnabla\btimes \left ( u_{\|e}\bhatz \right ) \Big \},\label{eq:divPI_perp}
\end{eqnarray}
where an expression involving a bracket with a scalar as one argument and a vector
as the other one stands for the vector whose components are
obtained as the brackets of the scalar with each components of the vector.

The dominant contribution to $\displaystyle{ ( \bnabla\bcdot
  {\boldsymbol \Pi}^{(1,u)}_e)_\perp}$, which enters 
Eq. (\ref{eq:drift1}), originates from the
second term on the rhs and writes  $(\delta^2/2)\bnabla \omega_{ze}$.
It is $O(\beta_e\varepsilon)$ for both types of wave. For the KAWs, 
the other terms are $O(\beta_e^{1/2}\delta\varepsilon^2)$, or
$O(\beta_e^{3/2}\delta\varepsilon^2)$ for the term involving $B_z$,
while for the WWs, they are all $O(\beta^{3/2} \delta \varepsilon^2)$.

\subsubsection{Heat flux contributions}
Let us now consider the contributions involving
$\displaystyle{\boldsymbol{\Pi}^{(1,q)}_e}$. 
It is easy to find that
\begin{equation}
  \bhatb\bcdot \bnabla\bcdot {\boldsymbol
    \Pi}^{(1,q)}_e=-\delta^2[B_z,q_{\perp e}]  \label{bdivPi2}
\end{equation}
and 
\begin{equation}
  (\bnabla\bcdot {\boldsymbol \Pi}^{(1,q)}_e)_\perp=\delta^2 \left \{
  -\frac{1}{2} \bnabla_\perp \left (\frac{\beta_e}{2} q_{\perp e}
  u_{\| e} + \bhatz\bcdot(\bhatb_\perp \btimes \bnabla_\perp q_{\perp e}) \right
  )+\nabla_\| \bnabla\btimes \left ( q_{\perp e} \bhatz \right )
   \right \},\label{eq:divPI_perpq}
\end{equation}
where we have used that, to the required order,
$\displaystyle{\bhatz \btimes(\bnabla_\perp \btimes \bhatb_\perp) =-\Delta_\perp
  A_\| =-\frac{\beta_e}{2} u_{\| e}}$.

\subsection{Second-order contributions}
At second order, since terms of order $O(\varepsilon^3)$ or smaller
are discarded, we can write
\begin{eqnarray}
&&{\Pi}^{(2)}_{e,ij}=\frac{\delta^2}{4}\mathsfi{M}_{ijkl} \Big [
  \frac{D^{(e)}{\Pi}^{(1)}_{e,kl}}{Dt} 
+ \left
  ({\delta}_{kn}{\Pi}^{(1)}_{e,ml}+{\delta}_{ln}
  {\Pi}^{(1)}_{e,mk} \right)\partial_m {u}_{e,n} \Big ].\label{eq:press-tensor2}
\end{eqnarray}
In this formula, it is also sufficient to use for
$\boldsymbol{\Pi}^{(1)}$ (keeping only terms of order $O(\varepsilon)$)
\begin{eqnarray}
  && {\Pi}^{(1,1)}_{e, ij} =-\frac{\delta^2 }{4} \Big [ {\epsilon}_{i3q} 
    \left (\partial_q {u^*}_{e,j} + \partial_j {u^*}_{e,q}
    \right ) +3{\epsilon}_{i3q} \partial_q
  {u^*}_{\| e}{\delta}_{j3}   \Big ]^{S},
\end{eqnarray}
where $\boldsymbol{u}^*_e=\boldsymbol{u}_e+q_{\perp e} \bhatz$, 
while $\mathsfbi{M}$ reduces to
$\displaystyle{\mathsfi{M}_{ijkl}= ({\delta}_{ik} + 3
  {\delta}_{i3}{\delta}_{k3}){\epsilon}_{jl3}+({\delta}_{jk} + 3
  {\delta}_{j3}{\delta}_{k3}){\epsilon}_{il3}
  }$.

We shall now calculate the second-order contribution for the
perpendicular components of the gyroviscous force
$(\nabla_j{\Pi}^{(2)}_{e, ij})_\perp $ by
considering all terms separately.

In the second term on the rhs of Eq. (\ref{eq:press-tensor2}),  we find by
inspection that
$m\neq 3$, $j\neq 3$ (the terms involving parallel
derivatives would be too small), $l\neq 3$ (as implied by the symmetries
of $\mathsfi{M}_{ijkl}$) and $k\neq 3$ (because $k=3$ only
contributes to the parallel gyroviscous force). It is then easy to
see that ${\Pi}^{(1,1)}_{e, i\neq 3, j\neq 3}$ only
involves perpendicular components of the velocity.
As a result, the order of magnitude of the considered term is
$\delta^4 \varepsilon^2/\mu^3=\beta_e^{3/2}\delta\varepsilon^2$ for
both types of wave, which is too small to be retained.
Similar arguments apply to the third and the first terms which also
contribute, at most, at order $\beta_e^{3/2}\delta\varepsilon^2$.
We thus conclude that $\displaystyle{  (\bnabla\bcdot {\boldsymbol
    \Pi}^{(2)}_e)_\perp =0}$.

Let us then turn to the parallel component $\bhatb\bcdot \bnabla
\bcdot \boldsymbol{\Pi}^{(2)}_{e} $ which reduces to $\partial_j
{\Pi}^{(2)}_{e, 3j}$. We have
$\displaystyle{\mathsfi{M}_{3jkl}=4 
  {\delta}_{k3}{\epsilon}_{jl3}}$ and
${\Pi}^{(1,1)}_{e,
  3j}=\delta^2{\epsilon}_{jq3}\partial_q u^*_{\|
  e}=\delta^2\bnabla\btimes (u^*_{\|e}\bhatz)$.
Up to terms of order $O(\epsilon^3)$, the first term of the rhs of Eq. (\ref{eq:press-tensor2}) writes
\begin{equation}
R_1=\delta^4
{\epsilon}_{jl3}\partial_j\frac{D^{(e)}}{Dt}({\epsilon}_{lq3}\partial_q
u^*_{\| e}) =-\delta^4 \left (\frac{D^{(e)}}{Dt}\Delta_\perp (u_{\|
  e}+q_{\perp e})
+\partial_j\boldsymbol{u}_{e \perp}\bcdot\bnabla_\perp \partial_j
(u_{\| e} +q_{\perp e})\right ),
\end{equation}
the second one writes
\begin{eqnarray}
  R_2&=&-\frac{\delta^4}{4}{\epsilon}_{jl3}\partial_j \left \{ \epsilon_{m3q}\left (
  \partial_q u^*_{e,l}+\partial_l u^*_{e,q}\right )\partial_m u_{\| e}+\epsilon_{l3q}\left (
 \partial_q u^*_{e,m}+\partial_m u^*_{e,q}\right )\partial_m u_{\| e} \right
 \}\nonumber\\
 &=&-\frac{\delta^4}{4}\left ( [\omega_{ze},u_{\| e}]+2
     \epsilon_{jl3}[u_{e,l},\partial_j u_{
     \| e}] +\Delta_\perp\boldsymbol{u}_{e\perp}\bcdot\bnabla_\perp u_{\| e} +
 2\partial_j\boldsymbol{u}_{e\perp}\bcdot\bnabla_\perp \partial_j u_{\| e} \right ),
 \nonumber\\
\end{eqnarray}
and the third one
\begin{equation}
R_3=\delta^4\left ([\omega_{ze},u^*_{\| e}]+\epsilon_{jl3} [u_{e,l},\partial_j u^*_{e
     \|}] \right ).
\end{equation}

We thus have
\begin{eqnarray}
  \bhatb\bcdot \bnabla \bcdot \boldsymbol{\Pi}^{(2)}_{e} &=&
  -\delta^4 \frac{D^{(e)}}{Dt}\Delta_\perp u_{\| e} -\frac{3\delta^4}{2}
  \partial_j\boldsymbol{u}_{e\perp}\bcdot\bnabla_\perp \partial_j u_{\| e}
  -\frac{\delta^4}{4} \Delta_\perp\boldsymbol{u}_{e\perp}\bcdot\bnabla_\perp u_{\| e}\nonumber\\
  &&+\frac{3\delta^4}{4}[\omega_{ze},u_{\| e}]+\frac{\delta^4}{2}
  \epsilon_{jl3} [u_{e,l},\partial_j u_{ \| e}] \nonumber\\
&&-\delta^4 \frac{D^{(e)}}{Dt}\Delta_\perp q_{\perp e} -\delta^4
  \partial_j\boldsymbol{u}_{e\perp}\bcdot\bnabla_\perp \partial_j
  q_{\perp e}+\delta^4[\omega_{ze},q_{\perp e}]\nonumber\\
&&+\delta^4
  \epsilon_{jl3} [u_{e,l},\partial_j q_{ \perp e}] 
    \label{eq:divPI_par2_nophi}
\end{eqnarray}
or, when expressing $\boldsymbol{u}_{\perp e}$ in terms of
$\varphi^*$,
\begin{eqnarray}
 \bhatb\bcdot \bnabla \bcdot \boldsymbol{\Pi}^{(2)}_{e}&=&-\delta^4
 \frac{D^{(e)}}{Dt}\Delta_\perp u_{\| e}
 +\frac{\delta^4}{2}[\Delta_\perp\varphi^*,u_{\|
     e}]-\delta^4\sum_{j=x,y}[\partial_j\varphi^*,\partial_j u_{\|
     e}]\nonumber\\
&&-\delta^4 \frac{D^{(e)}}{Dt}\Delta_\perp q_{\perp e}
 +\delta^4[\Delta_\perp \varphi^*,q_{\perp  e}]
 .\label{eq:divPI_par2}
\end{eqnarray}

Since all the terms at this order in the expansion of
Eq. (\ref{eq:press-tensor}) are $O(\epsilon^2)$,
the higher orders cannot contribute to the gyroviscous force at
order $O(\epsilon^2)$.

\subsection{Curl of the gyroviscous force}
We shall now turn to  the expressions entering Eq. (\ref{induc}).
Since only first-order terms contribute to the perpendicular
gyroviscous force, one gets from Eqs. (\ref{eq:divPI_perp}) and (\ref{eq:divPI_perpq})
\begin{eqnarray}
&&{\widehat{\boldsymbol z}}\bcdot \bnabla \times (\bnabla \bcdot
{\boldsymbol \Pi}_e)=\delta^2 \Big\{ [p_{\perp e}-B_z,\omega_{ze}]+
\sum_{i=x,y}[\partial_i (p_{\perp e}-B_z),\partial_i\varphi^*]\nonumber\\
&&\qquad +\frac{1}{2}\Delta_\perp \left ( \nabla_\| u_{\| e}- \bnabla\bcdot
  \boldsymbol{u}_e \right ) -\nabla_\| \Delta_\perp u_{\| e} +
  \sum_{i=x,y} [\partial_i A_\|, \partial_i u_{\|e }]\nonumber\\
&&\qquad  -\nabla_\| \Delta_\perp q_{\perp e}
+ \sum_{i=x,y} [\partial_i A_\|, \partial_i q_{\perp e}]
  \Big\}
\end{eqnarray}
which, after some algebra, rewrites
\begin{eqnarray}
&&{\widehat{\boldsymbol z}}\bcdot \bnabla \times (\bnabla \bcdot
{\boldsymbol \Pi}_e)=\delta^2 \Big\{ [p_{\perp e}-B_z,\Delta_\perp
\varphi^*]+
\sum_{i=x,y}
  [\partial_i (p_{\perp e}-B_z),\partial_i\varphi^*]\nonumber\\
  &&-\frac{1}{2}[\Delta_\perp A_\|, u_{\| e}] -\frac{1}{2}\Delta_\perp  \bnabla\bcdot
  \boldsymbol{u}_e -\frac{1}{2}\nabla_\| \Delta_\perp u_{\| e}
  -\nabla_\| \Delta_\perp q_{\perp e}
+ \sum_{i=x,y} [\partial_i A_\|, \partial_i q_{\perp e}]\Big
  \}.\nonumber\\
  \end{eqnarray}
For the KAWs, all the terms in this equation (except that involving $B_z$) are
$O(\beta_e \epsilon^2)$, while those of Eq. (\ref{induc}) are $O(\epsilon^2)$. 
In this formula, the term $ \bnabla\bcdot   \boldsymbol{u}_e$ can be
replaced by $\displaystyle{-\frac{D^{(e)}}{Dt}}n$. Differently, for the
WWs, $ \bnabla\bcdot   \boldsymbol{u}_e$ is negligible and the other terms
are all $O(\beta_e^2 \varepsilon^2)$,  while those of Eq. (\ref{induc}) are $O(\beta_e\epsilon^2)$. 

Finally, the term ${\widehat{\boldsymbol z}}\bcdot (\bnabla n \times
(\bnabla \bcdot  {\mathsfbi \Pi}_e))$ simply writes at order
$O(\epsilon^2)$
\begin{equation}
  {\widehat{\boldsymbol z}}\bcdot (\bnabla n \times
(\bnabla \bcdot  {\mathsfbi \Pi}_e)) =\frac{\delta^2}{2} [n, \Delta_\perp \varphi^*], \label{DnxDPi1}
\end{equation}
which is $O(\beta \varepsilon^2)$ for the KAWs where  $n=-
(1/\tau)\varphi$ and negligible for the WWs.

\section{Relation with classical FLR formulations} \label{append:classical}

It is of interest to relate  the FLR pressure tensor formulas derived using a systematic
asymptotics in Appendix \ref{append:FLR-pressure}, to the classical 
formulations given in \citet{Hazeltine85} and \citet{Hsu86}.

Neglecting the heat flux contribution and using the identity
\begin{equation}
[p_{\perp e}, \boldsymbol{u}_{\perp e}]=\left (\bnabla_\perp p_{\perp e}
\bcdot \bnabla_\perp \right )\bnabla_\perp \varphi^* -\Delta_\perp \varphi^*
\bnabla_\perp p_{\perp e}, \label{eq:B1}
\end{equation}
it is easily checked that, up to the sign coming from the electron charge, 
Eqs. (\ref{eq:divPI_par}) and (\ref{eq:divPI_perp}) (which provide the leading-order
contribution) identify with
Eq. (4.133) of \citet{Hazeltine85}   once we perform the 
assumptions made in this paper, namely isothermal electrons ($p_{\perp e}=n$),
no longitudinal magnetic fluctuations  ($B_z=0$)
nor longitudinal variations ($\nabla_\| =0$), negligible compressibility of
the transverse electron flow
($\bnabla_\perp\bcdot \boldsymbol{u}_{\perp e}=0$).

We now turn to Eq. (24) of \citet{Hsu86} 
which, in comparison with that of \citet{Hazeltine85}, takes into
account parallel gradients, parallel flow and compressibility.
Defining the electron diamagnetic drift
$\displaystyle{\boldsymbol{V}_{De} =- \bhatb \btimes \bnabla
p_{\perp e}}$, and using Eq. (\ref{eq:B1}), we note that, at  order $O(\epsilon^2/\mu)$,
\begin{equation}
\frac{D^{(e)}}{Dt} \boldsymbol{V}_{De}=\Delta_\perp \varphi^* \bnabla_\perp p_{\perp e}-
\left (\bnabla_\perp p_{\perp e} \bcdot \bnabla_\perp \right )\bnabla_\perp \varphi^*
- \left (\bhatz \btimes \bnabla_\perp \right ) \frac{D^{(e)}}{Dt}p_{\perp e}. \label{eq:B2}
\end{equation}
Under the assumption of isothermal electrons, the last term is evaluated using the continuity
equation. Putting together Eqs. (\ref{eq:divPI_par}), (\ref{eq:divPI_perp}), (\ref{eq:B1}) and neglecting $B_z$,  we obtain to leading order 
\begin{eqnarray}
  \bnabla\bcdot \boldsymbol{\Pi}_e=&&\delta^2 \Big\{
  \left (\bnabla_\perp p_{\perp e} \bcdot \bnabla_\perp \right )\bnabla_\perp
  \varphi^* -\Delta_\perp \varphi^* \bnabla_\perp p_{\perp e}+
  \frac{1}{2} \bnabla_\perp \left ( p_{\perp e} \Delta_\perp \varphi^*
  \right )
  \nonumber\\
  &&+\nabla_\| \Delta_\perp \varphi^* \bhatz +\bhatz\cdot \left (
  \bnabla_\perp p_{\perp e}\btimes \bnabla_\perp u_{\| e} \right ) \bhatz
  -\frac{1}{2}\bhatz \btimes \bnabla_\perp \left (\bnabla \bcdot
  \boldsymbol{u}_e \right )\nonumber \\
  &&+\frac{1}{2}\bhatz \btimes \bnabla_\perp \left ( \nabla_\| u_{\|
    e} \right ) -\nabla_\| \left ( \bhatz \btimes \bnabla_\perp
  u_{\| e} \right )+\sum_{i=x,y}
  [\partial_i\varphi^*,\partial_i A_\|] \bhatz\Big \}.\label{eq:our-flr}
\end{eqnarray}
 Using Eq. (\ref{eq:B2}) and the further approximation of a constant magnetic field ($A_\|=0$), this expression rewrites, 
\begin{eqnarray}
  \bnabla\bcdot \boldsymbol{\Pi}_e=&&\delta^2 \Big\{
-\frac{D^{(e)}}{Dt} \boldsymbol{V}_{De} + \frac{1}{2} \bnabla_\perp \left (
p_{\perp e} \Delta_\perp \varphi^*   \right ) 
  \nonumber\\
  &&+\nabla_\| \Delta_\perp \varphi^* \bhatz -\bhatz\cdot \left (
   \bnabla_\perp u_{\| e} \btimes\bnabla_\perp p_{\perp e} \right ) \bhatz
  +\frac{1}{2}\bhatz \btimes \bnabla_\perp \left (\bnabla_\perp \bcdot
  \boldsymbol{u}_{\perp e} \right ) \Big \},
\end{eqnarray}
which identifies with Eq. (24) of \citet{Hsu86} (which refers to ions and which makes the same assumptions), once we make the
appropriate sign change for the electron charge (remembering that
the diamagnetic drift is here defined with the opposite sign).
Note that Eqs. (22) and (23a) of \citet{Hsu86} are recovered when using the formulas of Appendix  (\ref{append:FLR-pressure}) adapted to the ions and pushed to second order. Note that the term $\displaystyle{  (\bnabla\bcdot {\boldsymbol
	\Pi}^{(2)}_e)_\perp}$, which in the present paper,  has to be kept in the  analysis of \citet{Hsu86}.

\bibliographystyle{jpp}
\bibliography{biblio}

\end{document}